\documentclass[bibyear]{aa} 

\def\HH        {H$_{2}$}

\def\dmeth     {CH$_{2}$DOH}
\def\methd     {CH$_{3}$OD}
\def\ace       {(CH$_{3}$)$_2$CO}
\def\mf        {HCOOCH$_{3}$}
\def\de       {(CH$_{3}$)$_2$O}
\def\water     {H$_2$O}
\def\ec        {C$_2$H$_5$CN}
\def\amm       {NH$_3$}

\def\kms       {km~s$^{-1}$}
\def\cmm       {cm$^{-2}$}
\def\cmmm      {cm$^{-3}$}
\def\vlsr      {$V_{\rm LSR}$}

\def\jb        {Jy~beam$^{-1}$}
\def\mjb       {mJy~beam$^{-1}$}
\def\Eup       {$E_{\rm up}/k$}

\usepackage{graphicx}
\usepackage{natbib}
\usepackage{txfonts}
\usepackage{multirow}
\begin{document}
   \title{Acetone in Orion BN/KL
   \thanks{Based on observations carried out with the IRAM Plateau de Bure Interferometer. IRAM is supported by INSU/CNRS (France), MPG (Germany), and IGN (Spain).}
   }
   \subtitle{High-resolution maps of a special oxygen-bearing molecule}
   \author{T.-C. Peng\inst{1,2,3}
          \and D. Despois\inst{1,2}
          \and N. Brouillet\inst{1,2}
          \and A. Baudry\inst{1,2}
          \and C. Favre\inst{4}
          \and A. Remijan\inst{5}
          \and A. Wootten\inst{5}
          \and T. L. Wilson\inst{6}
          \and F. Combes\inst{7}
          \and G. Wlodarczak\inst{8}
          }
   \institute{Univ. Bordeaux, LAB, UMR 5804, F-33270, Floirac, France
        \and CNRS, LAB, UMR 5804, F-33270, Floirac, France
        \and European Southern Observatory, Karl-Schwarzschild-Stra{\ss}e 2, D-85748, Garching, Germany \\
        \email{tpeng@eso.org}
        \and Department of Astronomy, University of Michigan, 500 Church St., Ann Arbor, MI 48109, USA 
        \and National Radio Astronomy Observatory, 520 Edgemont Road, Charlottesville, VA 22903-2475, USA 
        \and Naval Research Laboratory, Code 7213, Washington, DC 20375, USA 
        \and Observatoire de Paris, LERMA, CNRS, 61 Av. de l'Observatoire, 75014 Paris, France
        \and Laboratoire de Physique des Lasers, Atomes et Mol\'ecules, Universit\'e de Lille1, UMR 8523, 59655 Villeneuve d'Ascq Cedex, France
             }

   \date{}

 
  \abstract
   {}
   {As one of the prime targets of interstellar chemistry study, Orion BN/KL clearly shows different molecular distributions between large nitrogen- (e.g., \ec) and oxygen-bearing (e.g., \mf) molecules. However, acetone \ace, a special complex O-bearing molecule, has been shown to have a very different distribution from other typical O-bearing molecules in the BN/KL region. Therefore, it is worth investigating acetone in detail at high angular resolutions. 
   }
   {We searched for acetone within our IRAM Plateau de Bure Interferometer 3 mm and 1.3 mm data sets. Twenty-two acetone lines were searched within these data sets. 
   }
   {Nine of the acetone lines appear free of contamination. Three main acetone peaks (Ace-1, 2, and 3) are identified in Orion BN/KL. The new acetone source Ace-3 and the extended emission in the north of the hot core region have been found for the first time. An excitation temperature of about 150 K is determined toward Ace-1 and Ace-2, and the acetone column density is estimated to be 2--4$\times10^{16}$ \cmm\ with a relative abundance of 1--6$\times10^{-8}$ toward these two peaks. Acetone is a few times less abundant toward the hot core and Ace-3 compared with Ace-1 and Ace-2.}
   {We find that the overall distribution of acetone in BN/KL is similar to that of N-bearing molecules, e.g., \amm\ and \ec, and very different from those of large O-bearing molecules, e.g., \mf\ and \de. Our findings show the acetone distribution is more extended than in previous studies and does not originate only in those areas where both N-bearing and O-bearing species are present. Moreover, because the N-bearing molecules may be associated with shocked gas in Orion BN/KL, this suggests that the formation and/or destruction of acetone may involve ammonia or large N-bearing molecules in a shocked-gas environment.}


   \maketitle
%

\section{Introduction}

\begin{table*}
\caption{Observational parameters of the PdBI data sets}             
\label{table-data}      
\centering                          
\tabcolsep=0.11cm
\begin{tabular}{lcccccccc}        
\hline\hline                 
Bandwidth  & Observation date & Configuration & Flux conversion    & RMS noise         & $\theta_{\rm HPBW}$\tablefootmark{a} & $\delta{\rm v}$\tablefootmark{b}  & \multicolumn{2}{c}{$\theta_{\rm syn}$\tablefootmark{c}}       \\    
(GHz)      &                  &               & (1 \jb) & (\mjb) & (\arcsec)           & (\kms)            & (\arcsec$\times$\arcsec) & PA (\degr)              \\
       
\hline                        

101.178--101.717  &  2003--2006   &  BC  & 15.8 K  &   2.9 &  49.7  & 1.85  & $3.79\times1.99$  & 22 \\
203.331--203.403  &  10--11/1999  &  CD  &  7.0 K  & 170.1 &  24.2  & 0.92  & $2.94\times1.44$  & 27 \\
223.402--223.941  &  2003--2007   &  BC  & 17.3 K  &  21.2 &  22.5  & 0.84  & $1.79\times0.79$  & 14 \\
225.675--225.747  &  09--11/2005  &  D   &  2.9 K  &  40.8 &  22.3  & 0.42  & $3.63\times2.25$  & 12 \\
225.805--225.942  &  09--11/2005  &  D   &  2.9 K  &  40.8 &  22.3  & 0.42  & $3.63\times2.26$  & 12 \\
228.976--229.113  &  09--11/2005  &  D   &  1.8 K  &  43.9 &  21.5  & 0.42  & $6.02\times2.27$  & 10 \\

\hline                                   
\end{tabular}
\tablefoottext{a}{Primary beam size}
\tablefoottext{b}{Channel separation}
\tablefoottext{c}{Synthesized beam size with natural weightings}

\end{table*}


\begin{table*}
\caption{Spectral parameters of observed acetone transitions}             
\label{table-lines}      
\centering                          
\begin{tabular}{lcrrcc}        
\hline\hline                 
Frequency  & Transition                  & $E_{\rm up}/k $ & $S\mu^2$ & $g_{\rm s}$ & Comment\\    
(MHz)      & ($J_{K_{a},K_{c}}$)         & (K)           & (D$^2$)  &             &     \\
       
\hline                        

101297.450 (0.061) &$24_{13,12}-24_{12,13}$ EE & 235.5 &  105.7 & 16 & Partially blended with an NH$_2$CHO line \\ 
101426.664 (0.010) &      $9_{1,8}-8_{2,7}$ AE &  26.9 &   61.9 &  6 &   \\     
101426.759 (0.009) &      $9_{1,8}-8_{2,7}$ EA &  26.9 &   61.9 &  4 &   \\
101427.041 (0.010) &      $9_{2,8}-8_{1,7}$ AE &  26.9 &   61.9 &  2 &   \\
101427.130 (0.009) &      $9_{2,8}-8_{1,7}$ EA &  26.9 &   61.9 &  4 &   \\
101451.059 (0.008) &      $9_{1,8}-8_{2,7}$ EE &  26.8 &   61.9 & 16 &   \\
101451.446 (0.008) &      $9_{2,8}-8_{1,7}$ EE &  26.8 &   61.9 & 16 &   \\
101475.332 (0.011) &      $9_{1,8}-8_{2,7}$ AA &  26.7 &   61.8 & 10 &   Partially blended with an H$_2$CS line \\
101475.734 (0.011) &      $9_{2,8}-8_{1,7}$ AA &  26.7 &   61.8 &  6 &   Partially blended with an H$_2$CS line \\

203336.247 (0.036) &  $27_{4,23}-27_{3,24}$ EA & 225.8 &   39.0 &  4 &   Blended with an NH$_2$CHO line \\
	 	           &  $27_{5,23}-27_{4,24}$ EA & 225.8 &   39.0 &  4 &   Blended with an NH$_2$CHO line\\
203336.291 (0.037) &  $27_{4,23}-27_{3,24}$ AE & 225.8 &   13.0 &  6 &   Blended with an NH$_2$CHO line\\
	 	           &  $27_{5,23}-27_{4,24}$ AE & 225.8 &  117.1 &  2 &   Blended with an NH$_2$CHO line\\ 
  
223512.409 (0.012) &    $16_{8,9}-15_{7,8}$ EE & 103.8 &   60.9 & 16 &   Possibly blended with a C$_2$H$_5$$^{13}$CN line\tablefootmark{a} \\
223621.690 (0.016) &    $16_{8,9}-15_{7,8}$ AA & 103.7 &   60.9 & 10 &   Partially blended with an \mf\ line \\
223684.608 (0.012) &  $17_{6,11}-16_{7,10}$ EA & 110.8 &   80.6 &  4 &   Blended with the U223686.7 line\\
223684.610 (0.012) &  $17_{6,11}-16_{7,10}$ AE & 110.8 &   80.6 &  6 &   Blended with the U223686.7 line\\
223692.004 (0.012) &  $17_{7,11}-16_{6,10}$ EA & 110.8 &   80.6 &  4 &   Partially blended with the U223694.8 line\\
223692.104 (0.012) &  $17_{7,11}-16_{6,10}$ AE & 110.8 &   80.6 &  2 &   Partially blended with the U223694.8 line\\
223732.826 (0.032) &    $12_{9,4}-11_{8,4}$ EE &  66.1 &    8.7 & 16 &   Blended with the U223733.6 line \\
223767.585 (0.010) &  $17_{6,11}-16_{7,10}$ EE & 110.7 &   80.6 & 16 &   Blended with the U223767.6 line \\
223775.253 (0.010) &  $17_{7,11}-16_{6,10}$ EE & 110.7 &   80.6 & 16 &   Possibly blended with a $^{13}$C$_2$H$_5$CN line\tablefootmark{a} \\
223850.417 (0.014) &  $17_{6,11}-16_{7,10}$ AA & 110.7 &   80.5 & 10 &   Partially blended with an \mf\ line\\
223858.308 (0.145) &  $17_{7,11}-16_{6,10}$ AA & 110.7 &   80.5 &  6 &   Possibly blended with a $^{13}$C$_2$H$_5$CN line\tablefootmark{a}\\

225744.082 (0.011) &  $19_{4,15}-18_{5,14}$ EE & 122.1 &  117.6 & 16 &   \\
                   &  $19_{5,15}-18_{4,14}$ EE & 122.1 &  117.6 & 16 &   \\

225811.979 (0.046) &    $13_{9,5}-12_{8,4}$ EE &  75.4 &   36.9 & 16 &   Possibly blended with a $^{13}$C$_2$H$_5$CN line\\

228979.750 (0.023) &  $12_{10,2}-11_{9,2}$ EA  &  68.7 &   62.6 &  4 &   Blended with the U228979.0 line \\
229033.736 (0.015) &  $22_{1,21}-21_{2,20}$ AE & 133.1 &  173.2 &  2 &    \\
                   &  $22_{2,21}-21_{1,20}$ AE & 133.1 &  173.2 &  6 &    \\
229033.771 (0.015) &  $22_{1,21}-21_{1,20}$ EA & 133.1 &  173.2 &  4 &    \\
                   &  $22_{2,21}-21_{2,20}$ EA & 133.1 &  173.2 &  4 &    \\
229041.826 (0.024) &   $12_{10,3}-11_{9,3}$ EE &  68.6 &   61.1 & 16 &    \\
229055.797 (0.011) &  $22_{1,21}-21_{2,20}$ EE & 133.1 &   26.4 & 16 &    \\
                   &  $22_{1,21}-21_{1,20}$ EE & 133.1 &  146.8 & 16 &    \\
                   &  $22_{2,21}-21_{2,20}$ EE & 133.1 &  142.2 & 16 &    \\
                   &  $22_{2,21}-21_{1,20}$ EE & 133.1 &   31.0 & 16 &    \\
229058.049 (0.010) &    $14_{9,6}-13_{8,5}$ EE &  85.4 &   35.4 & 16 &   \\
229077.788 (0.018) &  $22_{1,21}-21_{2,20}$ AA & 133.0 &  173.2 &  6 &     \\ 
                   &  $22_{2,21}-21_{1,20}$ AA & 133.0 &  173.2 & 10 &    \\

\hline                                   
\end{tabular}

\tablefoot{Acetone spectroscopic data were taken from the JPL database based on the compilation of several data sets \protect\citep{Groner2002,Oldag1992,Vacherand1986,White1975,Peter1965}. Upper-state energy \Eup, effective line strength $S\mu^2$, and spin statistical weight $g_{\rm s}$ are also given. Additionally, only the transitions with an Upper-state energy below 300 K and a frequency uncertainty smaller than 1 MHz are listed here. Transitions with $E_{\rm up}/k$ higher than 300 K are all blended and not listed in this table.
\tablefoottext{a}{Possibly blended with $^{13}$C-substituted ethyl cyanide lines at HC, but their contributions are minor ($<$10\%, see also Fig. \ref{model-spectra-1}).}
}

\end{table*}


The Orion Molecular Cloud 1 (OMC-1) is a unique environment, especially in the surroundings of the Orion BN/KL region \citep{Becklin1967,Kleinmann1967}, and one of the closest massive star-forming regions \citep[413$\pm$10 pc,][]{Menten2007,Hirota2007,Sandstrom2007} for the study of interstellar chemistry. In this region, evidence of active star formation is overwhelming from detections of multiple strong infrared emission sources or embedded compact radio sources and strong or moderate molecular outflows. For example, a recent unusual explosive event \citep[see, e.g.,][]{Allen1993,Zapata2009} causes the finger-like shock structures seen in vibrationally excited \HH\ and high-velocity (30--100 \kms) CO emission. Although several sources in BN/KL have been considered as the origin of the explosive outflow, e.g., the continuum source SMA-1\footnote{SMA-1 is a submillimeter continuum peak detected by \citet{Beuther2004} located in the southwest of source $I$.} proposed by \citet{Beuther2008}, this explosion is very likely related to the disintegration of a stellar system, which involves the Orion BN object, Source $I$, and maybe Source $n$ \citep[see, e.g.,][]{Gomez2005,Zapata2009,Goddi2011a}. In addition, another outflow with a lower velocity (about 18 \kms) was detected through the proper motions of \water\ masers \citep[e.g.,][]{Genzel1981,Gaume1998} accompanied by SiO masers near the infrared source IRc2 \citep[e.g.,][]{Greenhill1998,Doeleman1999}. Evidence indicates that this low-velocity bipolar outflow along a NE-SW axis is driven by source $I$ \citep[see e.g.,][]{Plambeck2009}.

As one of the prime targets of interstellar chemistry study, Orion BN/KL clearly shows different molecular distributions between large nitrogen- (e.g., C$_2$H$_3$CN and C$_2$H$_5$CN) and oxygen-bearing (HCOOCH$_3$ and CH$_3$OCH$_3$) molecules \citep[see, e.g.,][and the references therein]{Blake1987,Liu2002,Friedel2008}. Since the critical densities are fairly similar for N- and O-bearing species, \citet{Friedel2008} have proposed the age effect, the time that the molecular gas has had to evolve from an initial composition after being released from ice mantles, may be the most critical factor in the Orion BN/KL N-O differentiation. In addition, they find that the distribution of acetone \ace\ is clearly different from those of typical N- or O-bearing molecules.

The differences between acetone and other large O-bearing molecules make acetone an excellent tool for testing different chemical models of complex molecules productions, i.e., in the gas phase by pure cosmic ray induced ion-molecule chemistry, on the surface of grains, or both. The higher temperatures in the vicinity of star-forming regions are certainly an important influence, because these allow some endothermic reactions to proceed. Another influence may be the recent shock event. If so, there would be a dependence on the time at which molecules were released to the gas phase and started to be processed by the gas-phase ion-molecule reactions.
 
   \begin{figure*}
   \centering
   \includegraphics[angle=-90,width=0.9\textwidth]{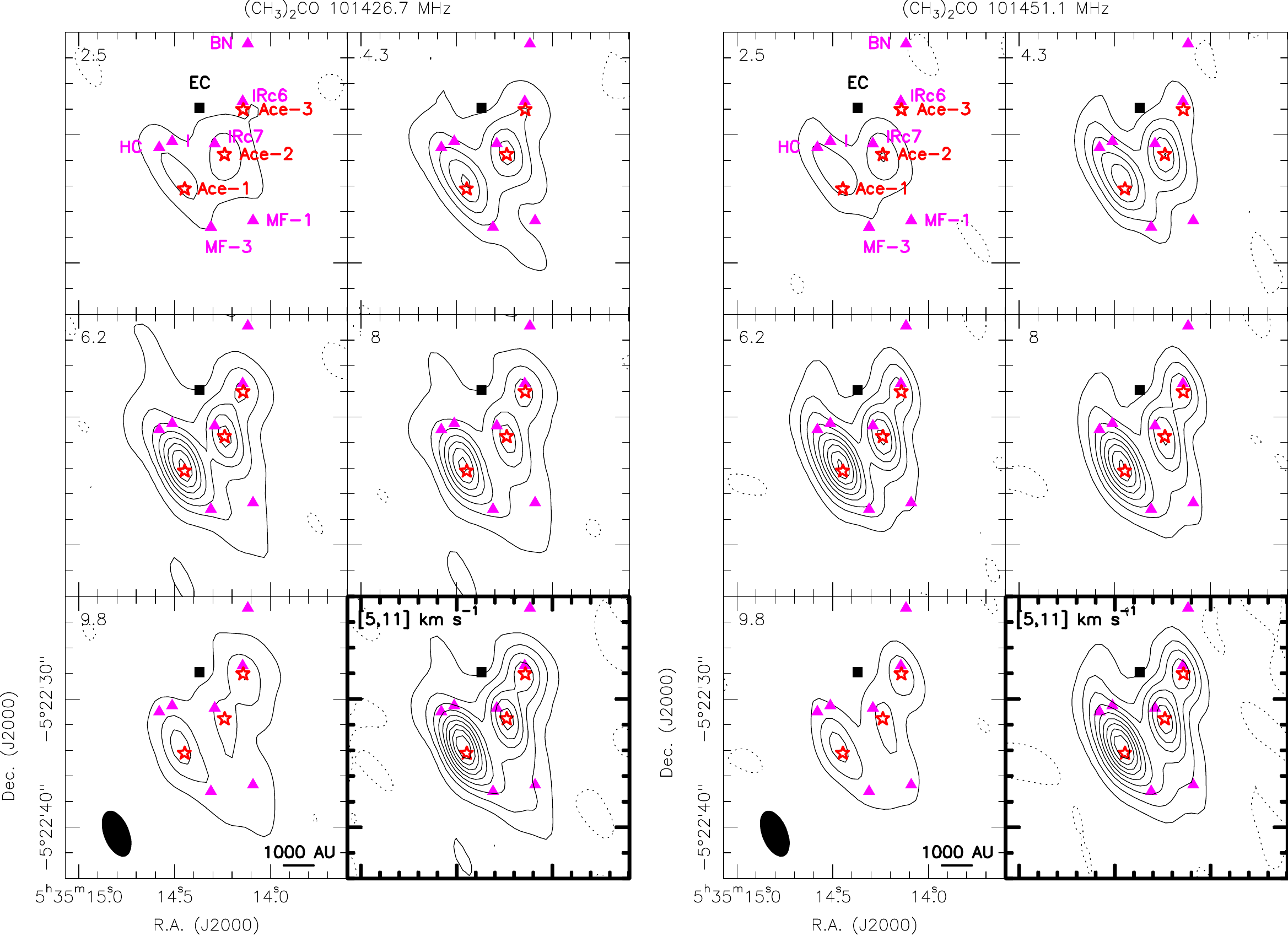}
   \caption{Left: Channel maps of the \ace\ emission at 101.43 GHz for a synthesized beam of $3\farcs8\times2\farcs0$, where four \ace\ lines, $9_{1,8}-8_{2,7}$ AE/EA and $9_{2,8}-8_{1,7}$ AE/EA (\Eup=26.9 K), are blended together. Contours run from 9 \mjb\ (3 $\sigma$) to 114 \mjb\ in steps of 15 \mjb, and the dashed contours represent --3 \mjb. The bottom-right panel shows the integrated intensity (from 5 to 11 \kms) in contours running from 10\% to 90\% in steps of 10\% of the peak intensity (623.1 \mjb\ \kms), and the dashed contours represent --5 \mjb\ \kms\ (1 $\sigma$). Right: Channel maps of the \ace\ emission at 101.45 GHz in a synthesized beam of $3\farcs8\times2\farcs0$, where two \ace\ lines, $9_{1,8}-8_{2,7}$ EE and $9_{2,8}-8_{1,7}$ EE (\Eup=26.8 K), are blended together. Contours run from 24 \mjb\ (8 $\sigma$) to 216 \mjb\ in steps of 24 \mjb, and the dashed contours represent --3 \mjb. The bottom-right panel shows the integrated intensity (from 5 to 11 \kms) in contours running from 10\% to 90\% in steps of 10\% of the peak intensity (1.1 \jb\ \kms), and the dashed contours represent --0.005 \mjb\ \kms\ (1 $\sigma$). The black square marks the center of explosion according to \citet{Zapata2009}. The positions of source BN, HC, infrared sources IRc6/7, source $I$, methyl formate peaks (MF-1 and MF-3) are marked as triangles. The positions of acetone emission peaks (Ace-1 to Ace-3) are marked as stars.}
              \label{Fig-ace-chmap-1}
    \end{figure*}

Acetone was first detected in the interstellar medium (ISM) by \citet{Combes1987} and later confirmed by \citet{Snyder2002} toward the hot molecular core Sagittarius B2(N-LHM). They derived an acetone column density of $2-9\times10^{16}$ \cmm\ with a rotational temperature of 100--270 K in a $\sim13\arcsec\times7\arcsec$ beam toward Sgr B2(N). The first acetone detection toward Orion BN/KL was reported by \citet{Friedel2005} using BIMA, and the follow-up observations at 1 mm with CARMA by \citet{Friedel2008} show that the acetone emission is compact in the Orion BN/KL region, and high angular resolution observations have the advantage of detecting weak signals from these complex organic species with compact spatial distributions. The derived acetone column density is $1-7\times10^{17}$ \cmm\ assuming a rotational temperature of 150--500 K toward the acetone emission peaks in BN/KL.

The formation process of acetone in space is still unclear. \citet{Herbst1990} have shown that the ion-molecule radiative association reaction $\rm{CH_3^++CH_3CHO\rightarrow(CH_3)_2CHO^+}$ proposed by \citet{Combes1987} is very inefficient even at low temperatures, where the rate coefficient for radiative association shows an inverse temperature dependence ($\approx T^{-1}$). Their results suggest that other reactions may in work, e.g., grain surface chemistry, because the radiative association reaction cannot account for the observed acetone abundance in Sgr B2. Therefore, the strategy for determining the most likely formation route of interstellar acetone involves a high angular and velocity comparison of the acetone spectrum with other large O-bearing molecules, and an attempt to correlate the distribution of the acetone emission where shocks or heating sources are likely to release molecules from icy grain mantles.

   \begin{figure*}
   \centering
   \includegraphics[angle=0,width=0.7\textwidth]{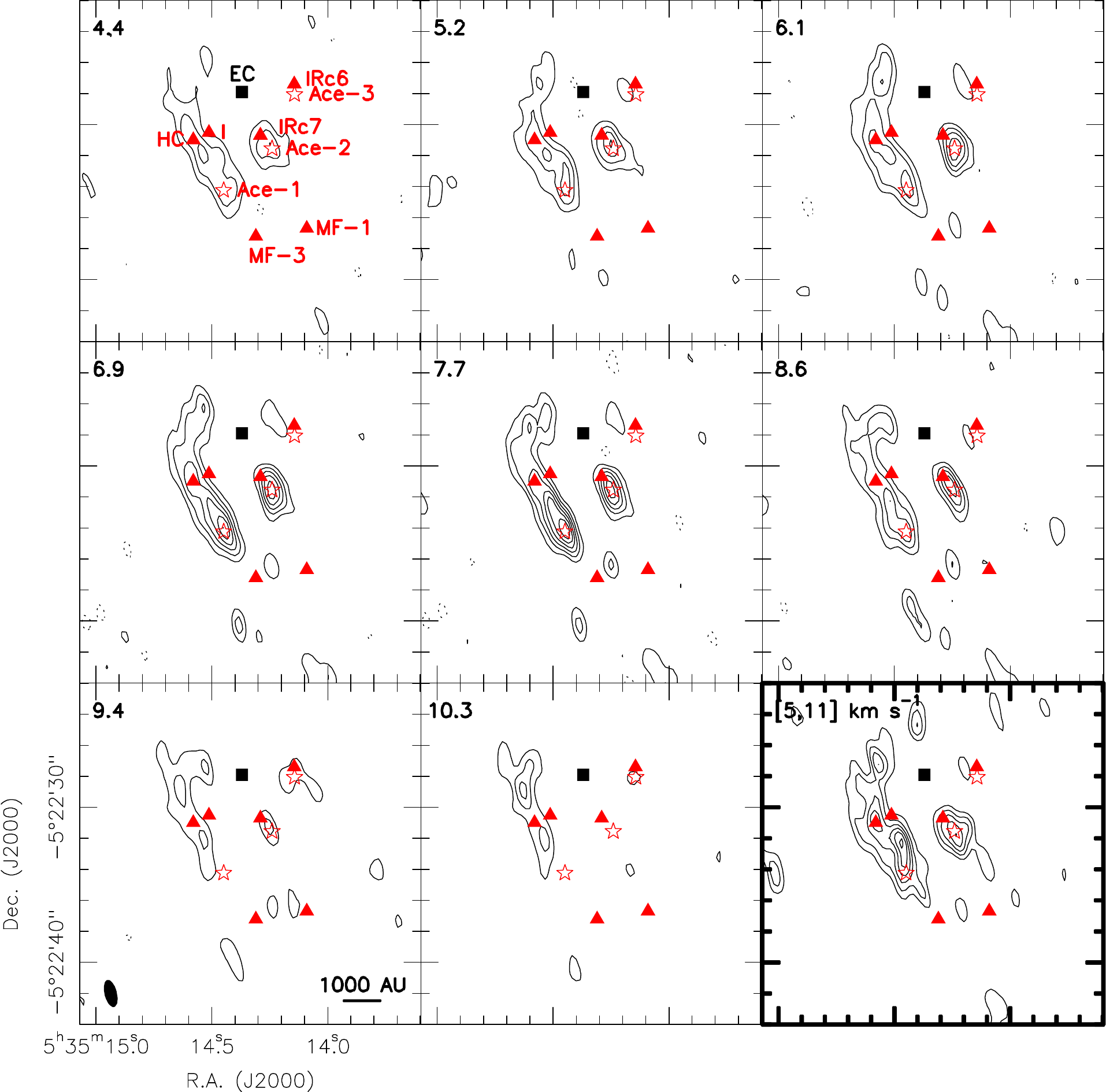}
   \caption{Channel maps of the \ace\ $17_{7,11}-16_{6,10}$ EE line (\Eup=110.7 K) at 223775.3 MHz for a synthesized beam of $1\farcs8\times0\farcs8$. Contours run from 40 \mjb\ (2 $\sigma$) to 320 \mjb\ in steps of 20 \mjb, and the dashed contours represent --20 \mjb. The bottom-right panel shows the integrated intensity (from 5 to 11 \kms) in contours running from 20\% to 95\% in steps of 15\% of the peak intensity (151.3 \mjb\ \kms). The black square marks the center of explosion according to \citet{Zapata2009}. The positions of source BN, HC, IRc6/7, source $I$, and methyl formate peaks (MF-1 and MF-3) are marked as triangles. The positions of acetone emission peaks (Ace-1 to Ace-3) are marked as stars.}
   \label{Fig-ace-chmap-2}
   \end{figure*}

The observations and acetone spectroscopy are presented in $\S$\ref{obs}, and the high-quality images of the acetone emission distribution at $<3$\arcsec\ resolution in Orion BN/KL are presented in $\S$\ref{results}. The spatial and spectral comparisons between acetone and other molecules are discussed in $\S$\ref{comparison}. Furthermore, the acetone formation and the cloud structure of BN/KL are discussed in $\S$\ref{discussion}. Moreover, we have compared our PdBI data with the science verification (SV) data of the Atacama Large Millimeter Array (ALMA) at 223 GHz in Appendix \ref{app2}. The three-dimensional visualization for different molecules in BN/KL is demonstrated in Appendix \ref{app3}.

\section{Observations and spectroscopy \label{obs}}

\subsection{IRAM observations and line blending}

The data used in this study are part of the large 1--3 mm data sets obtained from 1999 to 2007 using the PdBI\footnote{The IRAM Plateau de Bure Interferometer. IRAM is supported by INSU/CNRS (France), MPG (Germany), and IGN (Spain).} toward the Orion BN/KL region \citep[see][for more observational details]{Favre2011a}. Six data sets (Table \ref{table-data}) were used in this study, and the observations were carried out with five antennas equipped with two SIS receivers. The quasars 0458--020, 0528+134, 0605--085, and 0607--157, and the BL Lac source 0420--014 were observed for the phase and amplitude calibrations. The six units of the correlator allowed us to achieve different bandwidths and spectral resolutions. Using the IRAM 30m single-dish data (J. Cernicharo, priv. comm.), the missing flux for the \ace\ lines near 223 GHz is estimated as between 20\% and 50\%. The large uncertainty in this estimate is due to line confusion and the difficulty of determining spectral baselines in the 30m data. The missing flux in our \ace\ line interferometric observations near 101 GHz is estimated to be 25\% to 35\%.

The PdBI data were reduced with the GILDAS\footnote{http://www.iram.fr/IRAMFR/GILDAS/} package, and the continuum emission was subtracted by selecting line-free channels. Our continuum emission images were presented in \citet{Favre2011a} where the \HH\ column densities of selected clumps were also estimated. What we call continuum includes both true continuum emission (from the dust thermal and free-free emission) and a pseudo-continuum made of overlapping faint lines, which cannot be separated from the former true continuum. The data cube was then cleaned channel-by-channel with the Clark algorithm \citep{Clark1980} implemented in the GILDAS package.

     \begin{figure}
   \centering
   \includegraphics[angle=-90,width=0.40\textwidth]{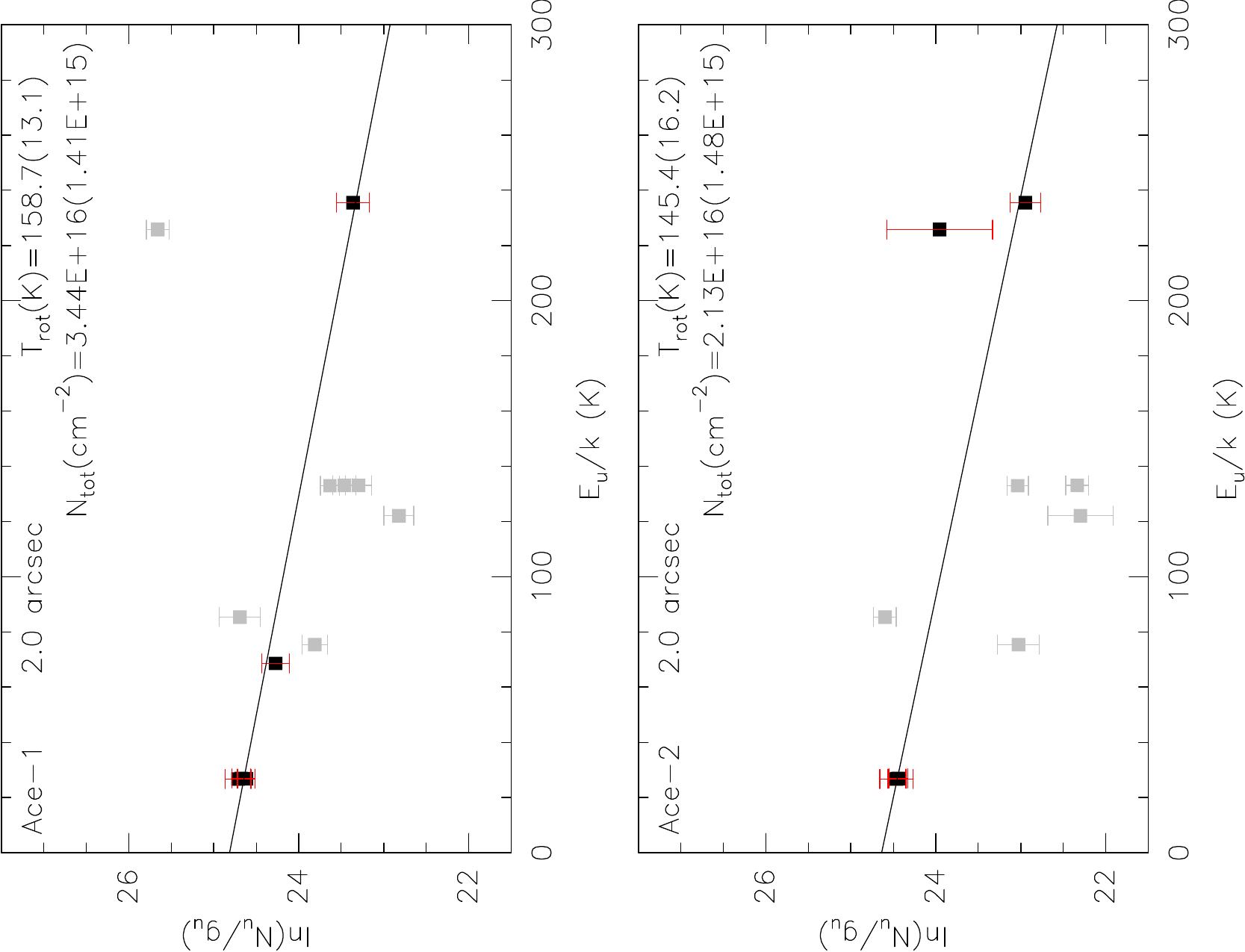}
   \caption{Acetone population diagrams toward Ace-1 and Ace-2. Only the lines without blending or slightly blended (can be decomposed) are used (black data points) in the fitting. The rest of the data points are shown in gray. The FWHM source size $\theta_{\rm s}$ is estimated to be 2 arcsec for Ace-1 and Ace-2 from the high-resolution acetone images. The beam filling factor $f=\theta_{\rm s}^2/(\theta_{\rm a}\theta_{\rm b}+\theta_{\rm s}^2$) is taken into account in the calculation, where $\theta_{\rm a}\theta_{\rm b}$ is the synthesized beam size.}
   \label{population-diagram-1}
   \end{figure}


Possible line blending with acetone was investigated by using the JPL database\footnote{http://spec.jpl.nasa.gov/}, the Cologne database for molecular spectroscopy (CDMS\footnote{http://www.astro.uni-koeln.de/cdms/}), and Splatalogue\footnote{http://www.splatalogue.net/}. It is known that $^{13}$C-substituted ethyl cyanide ($^{13}$CH$_3$CH$_2$CN, CH$_3^{13}$CH$_2$CN, and CH$_3$CH$_2^{13}$CN) has intense transitions in mm and submm regimes \citep{Demyk2007,Fortman2012}, and may blend with acetone lines in Orion BN/KL. We have examined the possible acetone line blending with $^{13}$C-substituted ethyl cyanide and found: (1) Their spatial distributions are concentrated at the hot core (HC, the strongest mm and submm continuum peak) region judging from the rather isolated CH$_3^{13}$CH$_2$CN $25_{5,20}-24_{5,19}$ line at 223426.5 MHz (not shown). (2) The synthesized spectrum of $^{13}$C-substituted ethyl cyanide indicates that their contributions to the acetone intensities are likely to be minor ($<$10\%) in our data (see Fig. \ref{model-spectra-1}) toward the main acetone emission peaks. The upper limit of $^{13}$C-substituted ethyl cyanide column density (beam averaged, $1\farcs8\times0\farcs8$) at Ace-1 is about $2\times10^{14}$ \cmm\ with a rotational temperature of 160 K (see also the synthesized spectra in Fig. \ref{model-spectra-1}). Additionally, a $^{13}$C-substituted ethyl cyanide column density of about $2-4\times10^{14}$ \cmm\ was estimated toward IRc2 (with a diameter of 7\arcsec) at 300 K, which is lower than the value ($1.6\times10^{15}$ \cmm\ with 300 K) given by \citet{Demyk2007} assuming the same source size toward IRc2. This difference may be because the $^{13}$C-substituted ethyl cyanide in Orion BN/KL is extended and was largely resolved out in our PdBI observations. Therefore, we conclude that the line blending due to $^{13}$C-substituted ethyl cyanide is not severe in our case.

        \begin{figure}
   \centering
   \includegraphics[angle=-90,width=0.40\textwidth]{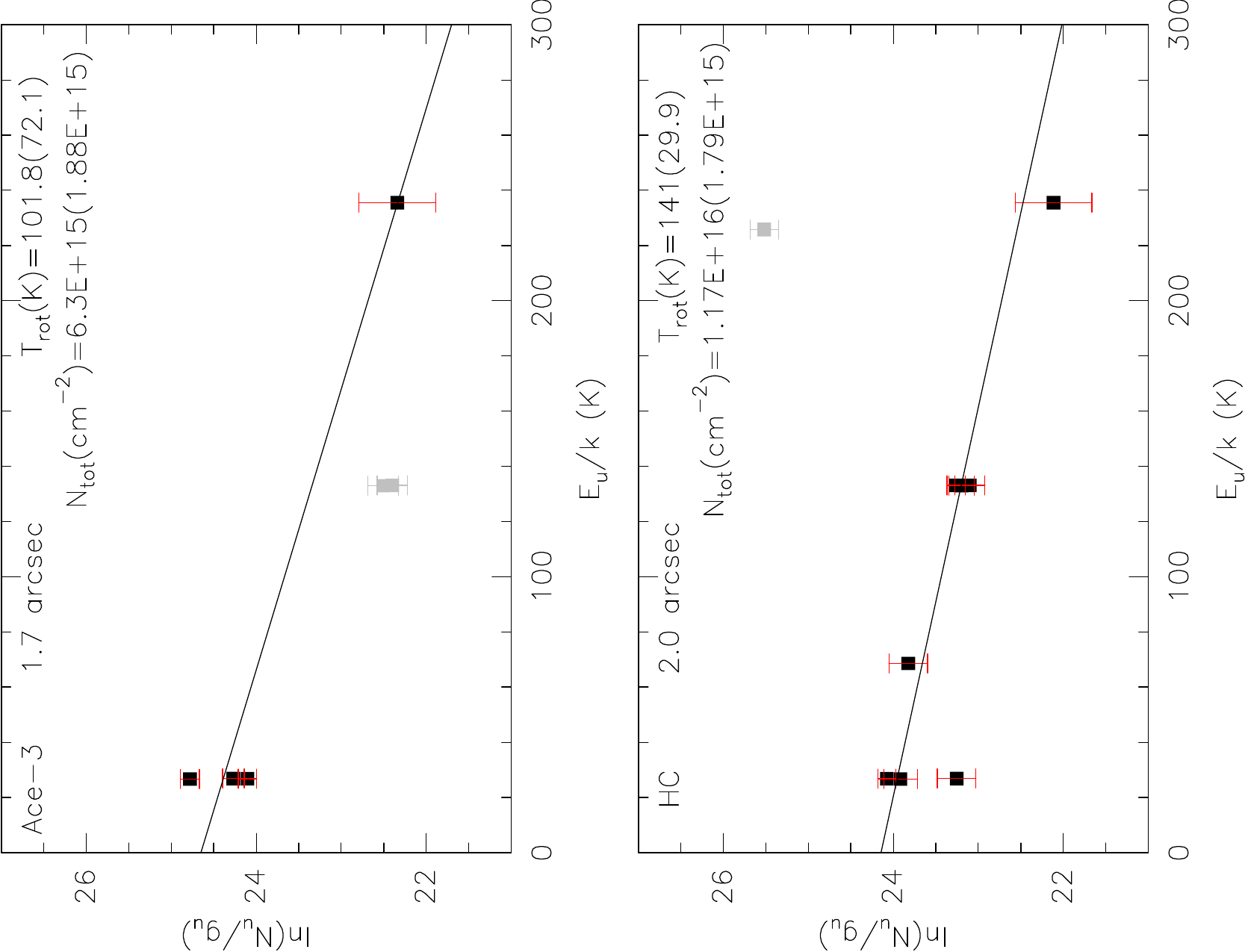}
   \caption{Acetone population diagrams toward Ace-3 and HC. Only the lines without blending or slightly blended (can be decomposed) are used (black data points) in the fitting. The rest of the data points are shown in gray. The FWHM source size $\theta_{\rm s}$ is estimated to be 1.7 arcsec for Ace-3 from the ALMA-SV images and 2 arcsec for Ace-2 from our PdBI images. The beam filling factor $f=\theta_{\rm s}^2/(\theta_{\rm a}\theta_{\rm b}+\theta_{\rm s}^2$) is taken into account in the calculation, where $\theta_{\rm a}\theta_{\rm b}$ is the synthesized beam size.}
   \label{population-diagram-2}
   \end{figure}

%

\subsection{\ace\ spectroscopy}

Acetone is a double methyl-substituted version of the ubiquitous species formaldehyde H$_2$CO. Owing to its two internal rotators (C$_{\rm 2V}$ symmetry), acetone has four torsional substates AA, EE, EA, and AE with slightly different rotational energy levels and different spin statistical weights \citep[see][for more details]{Groner2002}. The designations AE, EA, EE, and AA refer to the different symmetry states of the molecule. Under normal interstellar conditions, these different symmetry states will not interchange. Because of the different degenerate nature of the torsional substates, the spin weighting has to be taken into account in the column density calculation. The acetone spin weights for the $\rm{AA:EE:EA:AE}$ symmetry states are $6:16:4:2$ for $K_aK_c=ee, oo$ and $10:16:4:6$ for $K_aK_c=eo, oe$. The $b$-type asymmetric top selection rules ($\Delta J$=0, $\pm1$; $\Delta K_a$=$\pm1,3,...$; $\Delta K_c$=$\pm1,3,...$) apply to the transitions without severe torsional-rotational interactions \citep[see][]{Groner2002}. 

The effective rotational partition function $q$ is approximated by $q=261.67\ T^{1.5}$ for the temperature above 100 K \citep{Groner2002}. The rotational-vibrational partition function used by \citet{Friedel2005} is $Q_{\rm rv}\approx\sum\nolimits_{i=0}^2e^{-E_i/T}q$, where $E_i$ are the first three vibrational states at 0, 115, and 180 K \citep{Friedel2005}. However, we found that the acetone partition function given in the JPL database starts to deviate from the one used by \citet{Friedel2005} as $T\gtrsim120$ K (see Fig. \ref{partition}). It is likely that the approximation used by \citet{Friedel2005} does not take higher rotational and/or vibrational states into account which start to be essential at higher temperatures. Therefore, whereas the difference is limited for the temperature range ($\lesssim160$ K) derived here for acetone, one should use the more complete acetone partition function given in the JPL database for higher temperatures. The acetone molecular parameters in this paper were taken from the JPL database. 

\subsection{\ace\ critical density}

The critical densities $n_{\rm c}$ of acetone can be estimated according to
\begin{equation}
n_{\rm c}=\frac{\sum\nolimits_{l<u}A_{ul}}{\sum\nolimits_{l\neq u}C_{ul}},
\end{equation}
where $A_{ul}$ is the radiative de-excitation rate (Einstein A-coefficient) and $C_{ul}$ the collisional de-excitation rate \citep{Kwok2007}. Assuming a two-level system, the collisional de-excitation rate can be estimated as \citep[see][]{Friedel2008,Tielens2005,Flower2007}
\begin{equation}
C_{ul}=\left(\frac{8kT}{M_{\mu}\pi}\right)^{\frac{1}{2}}\sigma_{ul},
\end{equation}
where $M_{\mu}$ is the reduced mass of acetone and its main collisional partner \HH, and $\sigma_{ul}$ is the collisional cross section of acetone. Here we assume that the collisional cross section of acetone is approximated by its geometrical cross section ($\sim8\times10^{-16}$ \cmm). We find $C_{ul}\approx1.0\times10^{-10}$ cm$^3$~s$^{-1}$ for acetone at 150 K. The critical densities for the detected acetone transitions at 3 and 1.3 mm are estimated to be $4\times10^5$--$2\times10^8$ \cmmm. Therefore, the acetone emission generally probes a dense region with a density higher than a few times $10^5$ \cmmm\ at these frequencies, which is similar to \mf, \ec, and \de\ \citep{Favre2011a,Friedel2008}.

\section{Results \label{results}}

Our data include 22 acetone lines corresponding to 40 transitions (see Table \ref{table-lines}) with an upper energy state below 300 K. Nine of these lines appear free of contamination by other molecular emission. Three main acetone emission peaks (Ace-1, 2, and 3) were identified in the lower angular resolution images around 101.4 GHz. The acetone emission toward Ace-3 has been detected for the first time. The acetone transition parameters are summarized in Table \ref{table-lines}, and the line measurements toward selected positions are listed in Tables \ref{table-ace1} to \ref{table-hc}. The mean local standard of rest (LSR) velocities and FWHM line widths toward selected positions are listed in Table \ref{table-summary}. 
 

The LSR velocities of the selected positions are consistent with those reported by \citet{Favre2011a} for methyl formate and \citet{Chandler1997} for CS $J$=1--0. The average FWHM line width of acetone is about 3.5 \kms, similar to those of \mf\ and \de\ lines \citep{Favre2011a,Brouillet2013}. It is interesting to note that the acetone line widths of Ace-1 are narrower than those of Ace-2, Ace-3, and HC.

\subsection{Overall acetone distribution}

The acetone emission in BN/KL appears mainly at three peaks (Ace-1, Ace-2, and Ace-3) in our lower resolution images ($3\farcs8\times2\farcs0$, Fig. \ref{Fig-ace-chmap-1}), where the acetone emission around Ace-1 seems more extended than those around Ace-2 and Ace-3. The positions of the Ace-1 and Ace-2 peaks were determined at the central velocity channels of our highest resolution images ($1\farcs8\times0\farcs8$), and the Ace-3 position was determined on the lower resolution images due to the interferometric filtering effects and lower sensitivities of high-resolution images toward this position. The positions of the three acetone peaks are also consistent with the ALMA-SV data (Fig. \ref{pdbi-alma-chmap}). The arc-like structure in the north of HC is revealed for the first time in our highest resolution image (Fig. \ref{Fig-ace-chmap-2}). The reason \citet{Friedel2008} did not detect the arc-like extended emission and the Ace-3 peak is likely the different uv plane coverage and lower sensitivities. 

Ace-1 corresponds to the strongest methanol and deuterated methanol emission peak \citep[dM-1,][]{Peng2012a} and the second strongest methyl formate emission peak \citep[MF-2,][]{Favre2011a}. Ace-2 is close to the infrared source IRc7 \citep[see, e.g.,][]{Shuping2004,Gezari1998} but located about 1\arcsec\ southwest of its emission peak. In addition, Ace-3 is located at the same position of the MF-5 \citep{Favre2011a} and methanol (including deuterated methanol) peak KL-W \citep[e.g.,][]{Peng2012a}, and close to the infrared source IRc6 \citep[see, e.g.,][]{Shuping2004,Gezari1998}. The Ace-3 peak shows weak emission in our highest resolution images near 223 GHz because the extended and weaker acetone emission at Ace-3 is likely to be filtered out by the PdBI. We have also investigated the ALMA-SV data at the same frequency with slightly lower resolution and found clear detections of acetone toward Ace-3 (see Appendix \ref{app2} for more details). It indicates that the two velocity components seen in the acetone and methyl formate emission are most likely caused by the similar filtering effects. Alternatively, there is a possibility of two large-scale structures associated with Ace-3, which would be too extended to be detected by the PdBI \citep[see also][]{Chandler1997}. These large structures may extend to the outer part of the BN/KL region, which may also explain the two velocity components at about 6 and 11 \kms\ of the {\it Herschel}/HIFI O$_2$ detection \citep{Goldsmith2011} toward \HH\ Peak 1, about 26\arcsec\ to the northwest of Ace-3.

Acetone shows weak emission in the southern part of Orion BN/KL close to MF-1 and MF-3. However, the emission is extended (e.g., in the low resolution maps Fig. \ref{Fig-ace-chmap-1}) and also suffers from filtering effects in our PdBI data (see Fig. \ref{pdbi-alma-spectra-3}).

\subsection{Excitation temperature and column density}

We use population diagrams to estimate the column density of \ace\ \citep[e.g.,][]{Goldsmith1999,Turner1991}

\begin{equation}
\ln\frac{N_{\rm up}}{g_{\rm up}}=\ln\frac{3k\int T_{\rm B}dV}{8\pi^3\nu S\mu^2 g_{\rm s} f}=\ln\frac{N}{Q}-\frac{E_{\rm up}}{kT_{\rm rot}},
\end{equation}
where $N$ is the total column density, $T_{\rm rot}$ the rotational temperature, $S$ the line strength, $\mu$ the dipole moment \citep[$\mu_{b}=2.93$ D,][]{Peter1965}, $g_{\rm s}$ the spin statistical weight, and $f$ the beam filling factor. The mean optical depth of the acetone lines is estimated to be 0.02--0.03 toward the selected clumps near 101 GHz and less than 0.06 near 223 GHz with excitation temperatures of 100--150 K, taking the dust continuum emission into account and assuming a 90\% continuum flux loss due to filtering. The \ace\ beam-averaged column densities and rotational temperatures are estimated from a least-squares fit in the population diagrams. The uncertainties shown in these diagrams reflect the statistical fitting error in data where thermal noise and 10\% calibration uncertainty dominate. There is no attempt to estimate an absolute error, and we note that estimating the beam filling factor is uncertain. In addition, the column density uncertainties owing to the missing flux are likely to be less than a factor of two (see Appendix \ref{app2} for the comparison with the ALMA-SV data). The estimated acetone column densities, rotational temperatures, and fractional abundances toward the selected positions are summarized in Table \ref{table-summary}. 

\subsubsection*{Ace-1}

Toward Ace-1 (Fig. \ref{population-diagram-1}), an acetone column density is estimated to be $3.4\pm0.1\times10^{16}$ \cmm\ with an excitation temperature of $159\pm13$ K, assuming a source size of 2\arcsec. An \HH\ column density of $2.4\pm0.2\times10^{24}$ \cmm\ is derived toward the same position by using the 1 mm continuum emission \citep{Favre2011a}, assuming a dust temperature of 150 K and an absorption coefficient $\kappa_{\rm \nu}=5\times10^{-4}$ cm$^2$~g$^{-1}$. Therefore, the acetone abundance toward Ace-1 is $1.4\pm0.4\times10^{-8}$ and agrees with the result ($0.34-1.84\times10^{-8}$) of \citet{Friedel2005}.

\subsubsection*{Ace-2} 

We derived an acetone rotational temperature of $145\pm16$ K and a column density of $2.1\pm0.1\times10^{16}$ \cmm\ toward Ace-2 (Fig. \ref{population-diagram-1}), similar to those of Ace-1. From the 1.3 mm continuum emission of \citet{Favre2011a}, we derived an \HH\ column density of $3.8\pm2.0\times10^{23}$ \cmm\ toward Ace-2. Owing to its weaker dust continuum emission, the acetone fractional abundance ($5.5\pm4.0\times10^{-8}$) of Ace-2 seems to be higher than that of Ace-1 but with a larger uncertainty. 

\subsubsection*{Ace-3}

The acetone rotational temperature of Ace-3 is estimated to be $\sim100$ K, and the acetone column density to be $6.3\pm1.9\times10^{15}$ \cmm\ (Fig. \ref{population-diagram-2}). From the 1.3 mm continuum emission of \citet{Favre2011a}, we derived an \HH\ column density of $1.3\pm0.3\times10^{24}$ \cmm\ toward Ace-3 with a dust temperature of 100 K. The acetone abundance toward Ace-3 is $4.8\pm1.8\times10^{-9}$.

\subsubsection*{HC}

The acetone rotational temperature of HC is estimated to be about 141 K and the column density is estimated to be $1.2\pm0.2\times10^{16}$ \cmm\ (Fig. \ref{population-diagram-2}). From the 1.3 mm continuum emission of \citet{Favre2011a}, we have derived an \HH\ column density of $1.3\pm0.3\times10^{24}$ \cmm\ toward HC with a dust temperature of 140 K. The acetone abundance toward HC is $2.1\pm0.4\times10^{-9}$.

      \begin{figure*}
   \centering
   \includegraphics[angle=-90,width=0.9\textwidth]{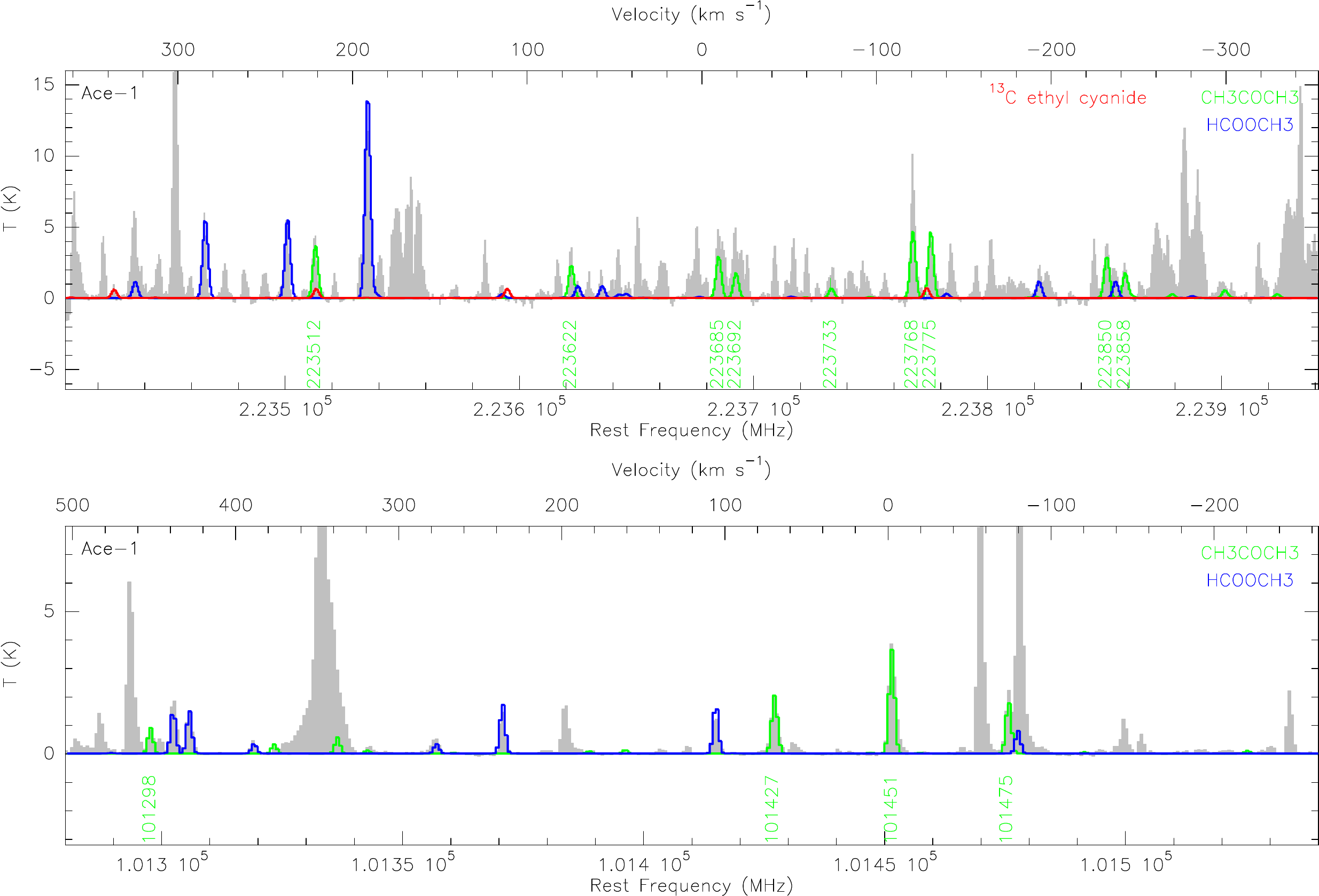}
   \caption{Comparison of the observed spectra (gray) and the synthesized acetone spectra (green) in LTE at 1.3 and 3 mm toward Ace-1. The observed spectra are plotted at their original synthesized beam sizes of $3\farcs8\times2\farcs0$ (101 GHz) and $1\farcs8\times0\farcs8$ (223 GHz). The acetone synthesized spectra near 101 GHz (upper panel) were produced by assuming a \ace\ column density of $3\times10^{16}$ \cmm\ and a temperature of 160 K, and the acetone synthesized spectra near 223 GHz (lower panel) were produced assuming a \ace\ column density of $4\times10^{16}$ \cmm\ with the same temperature. Line widths of acetone synthesized spectra were fixed at 3.5 \kms. As in Table \ref{table-lines}, only the transitions with $E_{\rm up}<300$ K are indicated with their frequencies (in MHz), and the others are all blended. Synthesized $^{13}$C-substituted \ec\ spectra are shown in red, assuming a column density of $2\times10^{14}$ \cmm\ and a temperature of 160 K. Their contributions to the acetone lines blending are minor. In addition, the methyl formate synthesized spectra (blue) are also shown for comparison.}
   \label{model-spectra-1}
   \end{figure*}

         \begin{figure}
   \centering
   \includegraphics[angle=-90,width=0.45\textwidth]{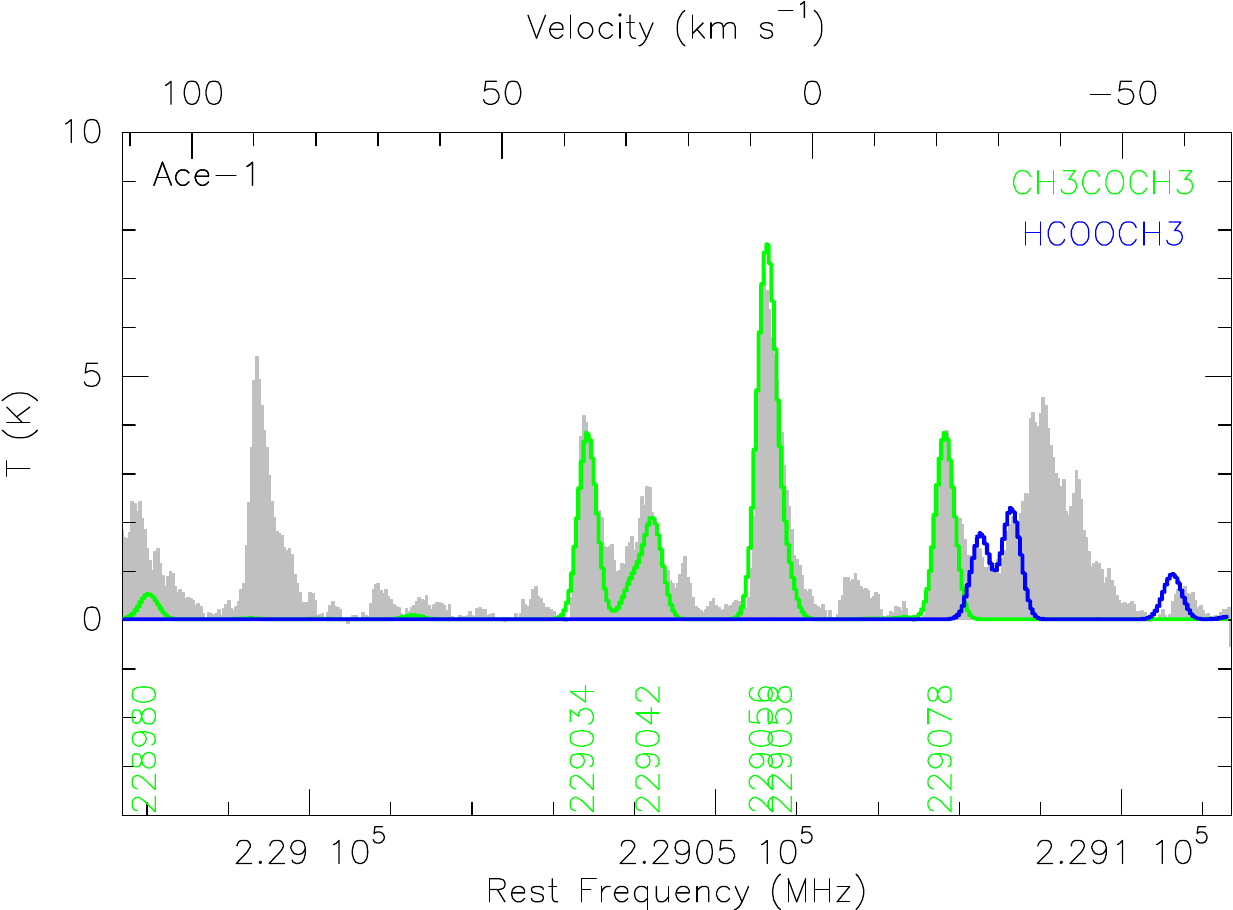}
   \caption{Comparison of the observed spectra (gray) and the synthesized acetone spectra (green) in LTE near 229 GHz toward Ace-1. The observed spectra are plotted at the original synthesized beam size of $6.0\times2.03$. The acetone synthesized spectra were produced assuming a \ace\ column density of $1.7\times10^{16}$ \cmm\ and a temperature of 160 K, and the line widths of acetone were fixed at 3.5 \kms. As in Table \ref{table-lines}, only the transitions with $E_{\rm up}<300$ K are indicated with their frequencies (in MHz), and the others are all blended. In addition, the methyl formate synthesized spectra (blue) are also shown for comparison.}
   \label{model-spectra-2}
   \end{figure}

\subsection{Synthesized spectra of acetone}

The synthesized spectra of acetone (Fig. \ref{model-spectra-1}) were produced with Weeds \citep{Maret2011} in the GILDAS package, assuming that the source sizes equal the PdBI synthesized beams at 101 and 223 GHz. Although the synthesized spectra do not fully reproduce the observed acetone intensities, the uncertainty of 10\%--20\% is likely contributed by line blending, different beam sizes, the accuracy of the acetone column density, excitation temperature, and possible non-LTE effects.

   \begin{figure*}
   \centering
   \includegraphics[angle=-90,width=0.8\textwidth]{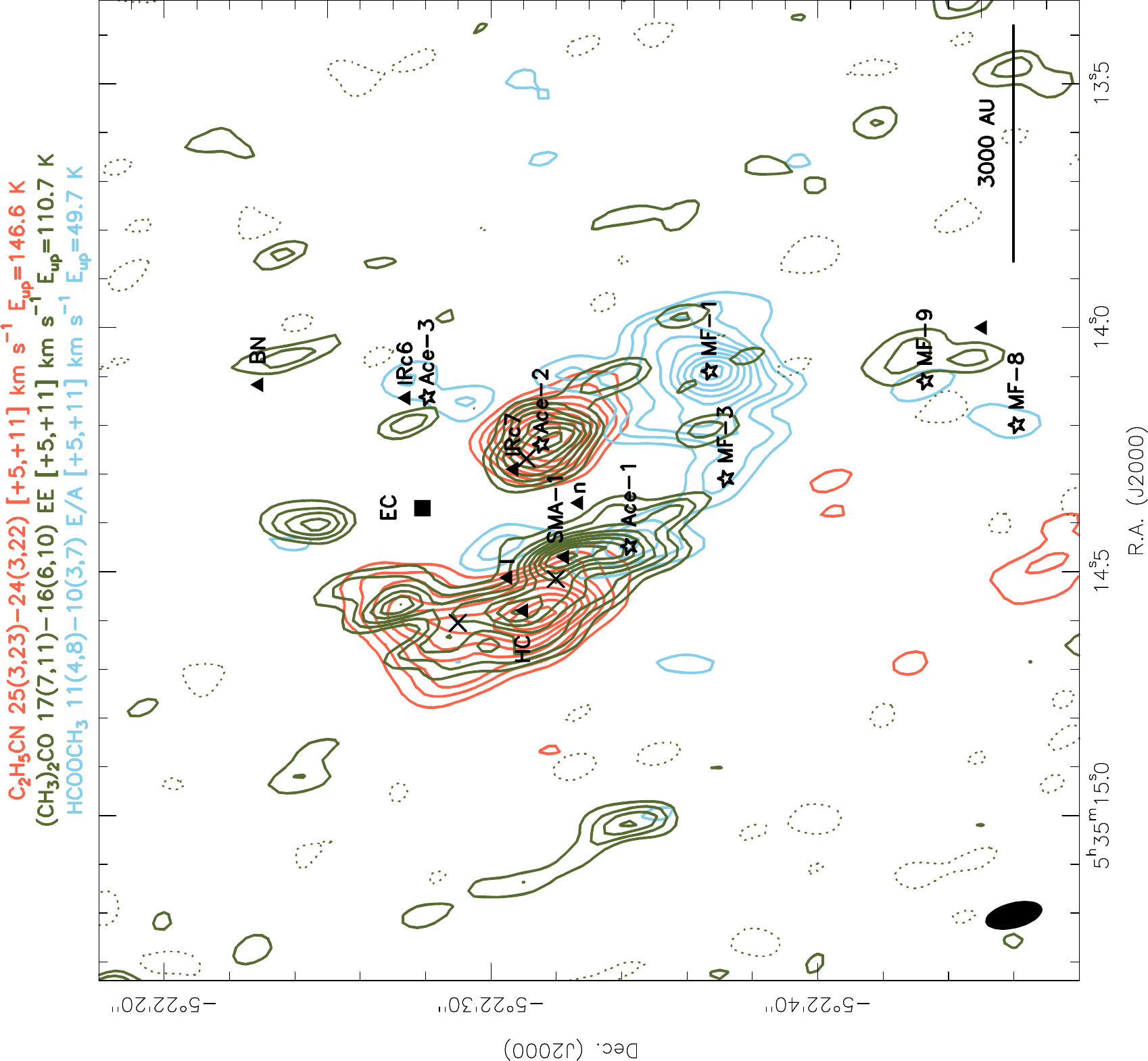}
   \caption{Spatial comparison of \ec, \mf, and \ace\ emissions with the same resolution ($1\farcs8\times0\farcs8$) in Orion BN/KL. The \ec\ $25_{3,23}-24_{3,22}$ line at 223553.6 MHz is shown in light-red contours. The \mf\ $11_{4,8}-10_{3,7}$ E/A lines at 223465.3 MHz and 223500.5 MHz are combined and shown in light-blue contours. The \ace\ $17_{7,11}-16_{6,10}$ EE line at 223775.3 MHz is shown in thick olive contours. Contours run from 15\% to 95\% in steps of 10\% of their peak intensities, and dashed contours represent $-10$\% of their peak intensities. The black square marks the center of explosion according to \citet{Zapata2009}. The positions of source BN, source $I$, source $n$, HC, SMA-1, and IRc6/7 are marked as triangles. The positions of acetone emission peaks (Ace-1 to Ace-3) and some of the methyl formate emission peaks (MF-1, MF-3, MF-8, and MF-9) are marked as stars. Black crosses mark the three NH$_3$ column density peaks (also close to the temperature peaks) according to \citet{Goddi2011b}.}
   \label{overlaid-1}
   \end{figure*}

   \begin{figure*}
   \centering
   \includegraphics[angle=-90,width=0.8\textwidth]{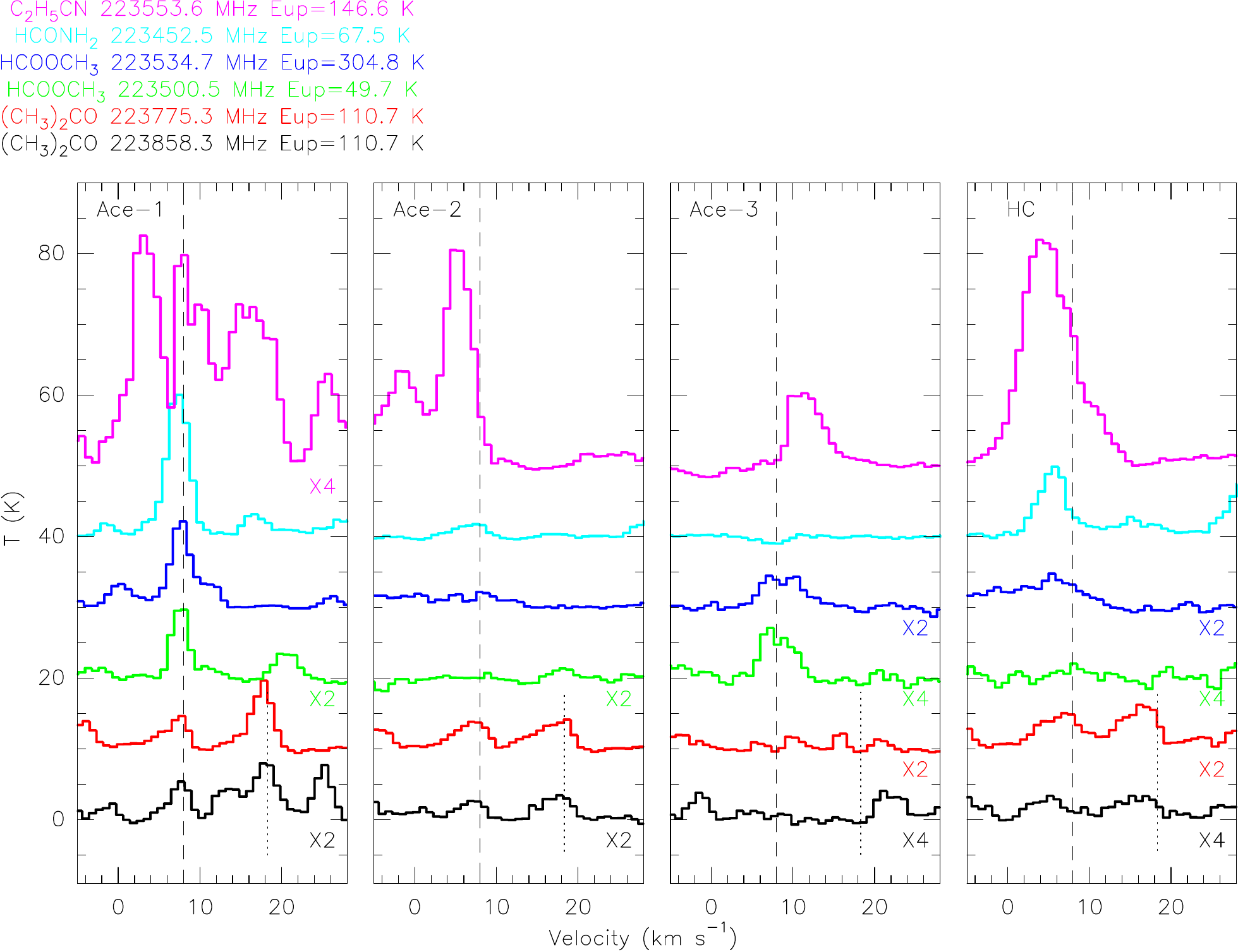}
   \caption{Comparison of the \ace, \mf, HCONH$_2$, and \ec\ spectra toward Ace-1, Ace-2, Ace-3, and HC with the same resolution ($1\farcs8\times0\farcs8$). Dashed lines indicate the \vlsr=8 \kms, and the dotted lines indicate the \ace\ lines at 223767.6 and 223850.4 MHz ($17_{6,11}-16_{7,10}$ EE/AA) in black and red, respectively. The spectra were shifted in intensity and some of them were multiplied for clarity. }
   \label{spectra-compare-1}
   \end{figure*}


\begin{table*}
\caption{Acetone line parameters toward selected positions}             
\label{table-summary}      
\centering                          
\tabcolsep=0.11cm
\begin{tabular}{lcccccccc}        
\hline\hline                 

Position  &  R.A.         & Dec.          & \vlsr\tablefootmark{a}  & $\Delta V$\tablefootmark{b}  & $T_{\rm rot}$\tablefootmark{d} & $N_{\rm tot}$\tablefootmark{d} & $X_{\rm ace}$\tablefootmark{d} & Note\tablefootmark{c} \\
          &  ($05^{\rm{h}}35^{\rm{m}}$)    & ($-05\degr22\arcmin$)     & (\kms)                  & (\kms)                       & (K)            & (\cmm)       &             &      \\
       
\hline                        

Ace-1     & $...14\fs449$ & $...34\farcs24$ & 7.47$\pm$1.07  & 3.49$\pm$1.09 & 158.7$\pm$13.1 & 3.4$\pm$0.1$\times10^{16}$ & 1.4$\pm$0.4$\times10^{-8}$ & dM-1, MF-2, CS-3\\
Ace-2     & $...14\fs240$ & $...31\farcs55$ & 7.58$\pm$1.27  & 4.46$\pm$1.30 & 145.4$\pm$16.2 & 2.1$\pm$0.1$\times10^{16}$ & 5.5$\pm$4.0$\times10^{-8}$ & CS-5, IRc7     \\
Ace-3     & $...14\fs143$ & $...28\farcs06$ & 6.70$\pm$1.95  & 4.20$\pm$1.94 & $\sim$100         & 6.3$\pm$1.9$\times10^{15}$ & 4.8$\pm$1.8$\times10^{-9}$ & MF-5, CS-6, IRc6, KL-W\\
HC        & $...14\fs580$ & $...31\farcs00$ & 6.87$\pm$1.29  & 4.92$\pm$1.37 & 141.0$\pm$20.9 & 1.2$\pm$0.2$\times10^{16}$ & 2.1$\pm$0.4$\times10^{-9}$ & CS-1\\

\hline                                   
\end{tabular}
\tablefoot{Positions are given at the epoch J2000.0. Acetone column densities and fractional abundances were estimated in source sizes of 2\arcsec\ for Ace-1, Ace-2, and HC, and 1\farcs7 for Ace-3.
\tablefoottext{a}{Mean LSR velocities measured at peak temperatures.}
\tablefoottext{b}{Mean FWHM line widths.}
\tablefoottext{c}{Counterparts as observed in the deuterated methanol (dM, Peng et al., 2012), methyl formate \citep[MF,][]{Favre2011a}, CS $J$=1--0 \citep{Chandler1997}, and infrared \citep[][]{Shuping2004,Gezari1998} emission. Note that IRc7 is offset $\sim1\arcsec$ to the southwest from Ace-2, and IRc6 is extended.}
\tablefoottext{d}{The acetone rotational temperatures $T_{\rm rot}$, column densities $N_{\rm tot}$, and fractional abundances ($X_{\rm ace}$=[\ace]/[\HH]) were estimated with source sizes of 2 arcsec.}
}

\end{table*}


\section{Comparison with other molecules \label{comparison}}

\subsection{Spatial comparison}

The detection of two molecules with strong and not blended lines, \ec\ and \mf, within the same data sets as acetone allows us to directly compare their distributions with the same observational parameters. In Figure \ref{overlaid-1}, the integrated intensity maps of \ec, \ace, and \mf\ were overlaid on the same velocity interval. The emission of the large N-bearing molecule \ec\ is mainly located in the northern part of the Orion BN/KL region, i.e., close to HC and IRc7, similar to the highly excited \amm\ \citep{Goddi2011b}. The emission of the typical O-bearing molecule \mf\ appears mostly in the south, i.e., the Compact Ridge region with a typical \vlsr\ of about 8 \kms, whereas the acetone emission appears in the regions where the \ec\ emission is also present. It is interesting to note that the \mf\ emission does not appear in the regions associated with hot ($\approx400$ K) and dense ammonia \citep{Goddi2011b}. Since the emissions of \ec\ and highly excited \amm\ (i.e., inversion transitions from $J, K\geq9, 9$) share a similar distribution, it is likely that \ec\ traces somewhat higher temperatures than \mf\ does, and \ace\ may be excited in the intermediate temperature range \citep[see also][]{Friedel2008}. For example, the excitation temperature of \ec\ toward the HC region is about 150 K, and is about 100 K for \mf\ toward the Compact Ridge region \citep{Favre2011a,Blake1987}. It is probably related to the acetone formation route in the interstellar medium (see \S\ref{discussion}). The high similarity of the acetone and \ec\ distributions is also seen in the 3D images in Appendix \ref{app3}, where both molecules show a similar arc-like structure that may be associated with the explosive event.

\citet{Chandler1997} obtained a high-resolution ($\theta_{\rm syn}=2\farcs1\times1\farcs7$) CS $J$=1--0 image with VLA, and they identified six main CS peaks in Orion BN/KL. It is essential to note that all acetone peaks have counterparts in the CS $J$=1--0 emission. The CS peaks CS-3, 5, and 6 correspond to Ace-1, 2, and 3, respectively, and CS-1 is located at the HC position (see Table \ref{table-summary} for the counterpart summary). In addition, the CS $J$=1--0 emission is strong in the south of BN/KL (i.e., CS-4) close to MF-1 and 3, where most of the large O-bearing molecules are detected. The overall CS distribution covers the region where large N- and O-bearing molecules are detected. The sulfur-bearing molecules (e.g., CS, SO, and SO$_2$) mainly form via warm gas-phase chemistry or shock chemistry \citep[e.g.,][]{Charnley1997,Pineau des Forets1993}, and they coincide with both N- and O-bearing molecules in BN/KL \citep{Schreyer1999,Chandler1997}. This suggests that the formation routes via warm gas phase or shock chemistry may lead to the formation of acetone in BN/KL.

Furthermore, \citet{Chandler1997} noticed that the integrated CS emission does not coincide with any identified clumps. This may be because the velocity dispersion in the arc-like structure is higher, and the integrated flux is dominated by the line width rather than the peak intensity. Although the acetone integrated intensity shown in Figures \ref{Fig-ace-chmap-2} and \ref{overlaid-1} peaks at SMA-1 \citep{Tang2010,Beuther2004}, it is likely to be caused by increased velocity width, similar to the CS emission peak mentioned above.

\subsection{Spectral comparison}

In Figure \ref{spectra-compare-1}, we compare the line profiles of \ace\ with \ec\ and \mf, together with formamide (HCONH$_2$), which is an interesting molecule because it contains one oxygen and one nitrogen atom and was detected in our data set near 223 GHz. It is interesting to note that the acetone spectra toward Ace-1 and Ace-2 show strong blueshifted line wings at about 4 \kms. The HCONH$_2$ $11_{1,11}-10_{1,10}$ spectrum toward Ace-2 also shows a blueshifted line wing. These blue-wing features may result from shocks. On the other hand, the \mf\ spectra do not have clear line wings. 

Toward Ace-3, the \ace\ and \mf\ spectra reveal that at least two velocity components are present. However, a velocity shift in \vlsr\ of the two components is observed, where the velocity difference between the two components is larger in acetone ($\sim4$ \kms) than in methyl formate ($\sim2-3$ \kms). In addition, the ALMA-SV acetone spectrum toward Ace-3 shows redshifted line wings (see Figure \ref{pdbi-alma-spectra-2}), which is not clear in our PdBI data. The differences observed between the ALMA-SV and PdBI data are most probably due to different uv coverages and sensitivities (see Table \ref{table-alma}). Because of its extended spatial nature, the acetone emission from Ace-3 is less likely to be recorded in our PdBI data.

Similarly, blueshifted wings can be seen in CS $J$=1--0 toward CS-3 (Ace-1) and CS-5 (Ace-2) \citep{Chandler1997}. Additionally, the CS $J$=1--0 line widths are broader in CS-6 (Ace-3) than in CS-3 and CS-5. The two velocity components seen in CS-6 are also due mainly to the filtering effects similar to acetone and methyl formate, compared with the ALMA-SV data. CS-3 (Ace-1) has a rather narrow line width of 3.3 \kms, similar to acetone. Therefore, most of the molecular lines in some regions of BN/KL (i.e., HC/CS-1, Ace-2/CS-5, and Ace-3/CS-6) show broader line widths than those in the other regions (i.e., Ace-1/CS-3, MF-3/dM-2/CS-4, and dM-3/MF-1) indicates that some regions may be more directly affected by shocks than others. These shocks could be driven by local outflow phenomena and/or the Orion explosion event.

\section{Discussion \label{discussion}}

      \begin{figure*}
   \centering
   \includegraphics[page=5,width=0.8\textwidth]{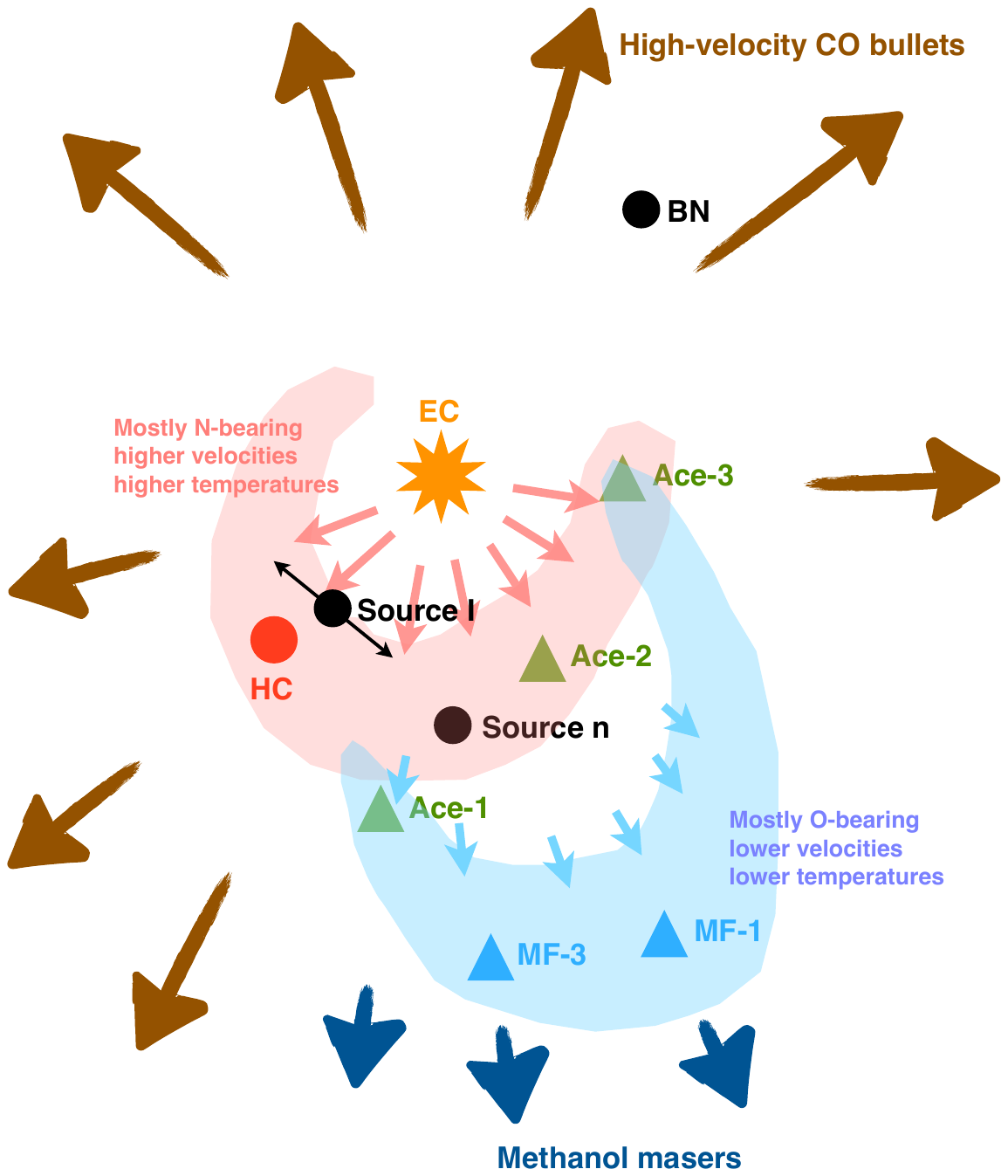}
   \caption{Cartoon presentation of the Orion BN/KL structure.}
   \label{cartoon}
   \end{figure*}

   \begin{table*}
\caption{Characteristics of different molecules in Orion BN/KL}             
\label{table-molecules}      
\centering                          
\tabcolsep=0.11cm
\begin{tabular}{lccccccccccc}        
\hline\hline                 

Species & Example & Distribution & \multicolumn{6}{c}{Main peak\tablefootmark{a}} & Feature\\
\hline                        
%
%

S-bearing  & CS                &  whole  &  CS-1 & CS-5  & CS-6  & CS-3  & CS-4\tablefootmark{b}  & CS-4\tablefootmark{b}        &  both N- and O-bearing features \\
N-bearing  & \amm, \ec\        &  north  &  HC   & IRc7  & KL-W  & ...   & ...   & ...         &  higher $T$ and broader $\Delta V$ \\
O-bearing  & \ace\             &  north  &  HC   & Ace-2 & Ace-3 & Ace-1 & ...   & ...         &  higher $T$ and broader $\Delta V$ \\
           & \mf, \de\         &  south  &  ...  & ...   & MF-5  & MF-2  & MF-1  & MF-3        &  lower $T$ and narrower $\Delta V$ \\
           & \dmeth, \methd\   &  south  &  ...  & ...   & KL-W  & dM-1  & dM-3  & dM-2        &  lower $T$ and narrower $\Delta V$ \\

\hline                                   
\end{tabular}
\tablefoot{
\tablefoottext{a}{The strongest peaks of different molecules. Orion KL-W is close to IRc6 and MF-1 is close to IRc4.}
\tablefoottext{b}{CS-4 is located between MF-1 and MF-3 but closer to MF-3.}
}
\end{table*}

\subsection{Acetone formation}

Some studies have suggested there is formation of carbonyl species, such as H$_2$CO, CH$_3$CHO, and \ace, on or within icy grain mantles \citep[e.g.,][]{Snyder2002,Bennett2005}. In the model of \citet{Garrod2008}, acetone is said to form by addition of CH$_3$ to CH$_3$CO and behaves similarly to \mf\ and \de. Recent results from \citet{Favre2011b} and \citet{Brouillet2013} show good agreement in the distributions between \mf\ and \de\ in BN/KL. However, acetone clearly has a different distribution from those of \mf\ and \de, and more similar ones to that of \ec. It is not clear whether the acetone binding energy (it evaporates from the ice mantle at about 65 K) used in the model of \citet{Garrod2008} is a critical factor or not. The acetone excitation temperature toward Ace-1 is estimated to be about 150 K, and it agrees with that of \mf\ \citep[140$\pm$14 K,][]{Favre2011a}. Therefore, \ace\ and \mf\ seem to trace the same gas toward Ace-1. On the other hand, not detecting \mf\ toward the Ace-2/IRc7 position indicates that different formation processes from \ace\ and \mf\ are involved. It is possible that some forming pathways of acetone have not been taken into account in the previous modelings. For example, \citet{Chen2011} carried out a theoretical investigation of possible reactions between carbonyl species and ammonia and show that acetone may react with water and ammonia via nucleophilic addition with low barriers. Although the final products of these reactions on icy grain mantles are not clear, further research is important. Because of its similar distribution to the N-bearing molecules, the formation of acetone may involve ammonia or N-bearing molecules or need similar conditions, e.g., shocks. 

It has been proposed by \citet{Remijan2004} that the formation of acetic acid (CH$_3$COOH) favors the hot molecular cores with well-mixed N- and O-bearing molecules. This may be another example that the formation of a large O-bearing molecule involves N-bearing molecules or needs similar conditions, although acetic acid has not been detected so far toward BN/KL \citep{Remijan2003}. However, acetone emission is detected toward the positions (i.e., IRc7 and arc-like structure in the north) where there is no strong emission of large O-bearing species like \mf\ and \de\ \citep{Favre2011a,Brouillet2013}. Furthermore, it is interesting to search for the molecules with distributions similar to acetone in Orion BN/KL in the future. Investigations of molecules related to acetone will be important for understanding the origins of interstellar acetone. 


\subsection{Overall cloud structure of BN/KL}

A clearer picture has emerged from the investigations of different molecules' distribution and the recent understanding of the stellar feedbacks inside the Orion BN/KL region. It has been shown that the explosive outflow that happened about 500 years ago may affect the BN/KL region in many ways. For example, \citet{Zapata2011} propose that HC, the strongest mm and submm continuum peak, is probably heated by the explosive outflow owing to its lack of embedded heating sources. In addition, \citet{Peng2012b} suggest that the high velocity bullets from the explosive outflow may be absorbed by the BN/KL cloud and induce the methanol masers seen in the south. The absorbed kinetic energy can then be transferred to thermal energy that affects the chemical processes in the cloud. As expected for the decelerated bullets, the thermal energy (or velocity) deposit is higher close to the explosion center than close to the stop point. This is clearly seen in the north-south differences in excitation temperatures and line widths of many molecules (see Table \ref{table-molecules}).

Figure \ref{cartoon} shows a cartoon overview of different molecular distributions in BN/KL. As discussed above, because the spectral features of CS are consistent with other molecules toward the same clump (e.g., smaller line widths in the south compared to the north of the cloud), the N-O differentiation seen in BN/KL is very likely caused by physical conditions (i.e., warmer and shocked gas) instead of their chemical properties (e.g., ice mantle composition). On the other hand, the age effect cannot be ruled out since different outflows in the northern region (e.g., the bipolar outflow of source $I$) may contribute, too, and the reason the northern part is more turbulent may be its more evolved nature. Nevertheless, the bipolar outflow of source $I$ alone cannot explain the blueshifted wings seen in most of the acetone peaks, because the outflow lies along a NW-SE axis \citep[see, e.g.,][]{Plambeck2009} and its orientation tends to be constant when moving from the explosion center to the present position \citep{Bally2011}. In addition, the acetone peaks are located in the direction of the source $I$ southern outflow lobe, which appears to be more redshifted than the northern one, but the line wings associated with Ace-1 and Ace-2 are blueshifted.

\section{Conclusions}

We have explored the acetone distribution in Orion BN/KL at angular resolutions between $1\farcs8\times0\farcs8$ and $6\farcs0\times2\farcs3$ (about 300--2500 AU) from the 3 mm and 1.3 mm Plateau de Bure interferometric data sets. The main findings follow.

\begin{enumerate}

\item Our sample consists of 22 lines of acetone with an upper energy state below 300 K, and nine of these lines appear free of contamination by other molecular emission. The FWHM line width of these acetone lines is about 3.5 \kms, similar to those of \mf\ and \de\ lines.

\item The acetone distribution exhibits three main peaks (Ace-1, 2, and 3) and the extended emission toward the Hot Core and the northern region, which appears to be arc-like. Ace-1 is also the strongest peak of the deuterated methanol emission \citep[dM-1,][]{Peng2012a} and second strongest peak of \mf\ \citep[MF-2][]{Favre2011a}. Ace-2 is located close to the peak of the IR source IRc7 with a separation of about 1\arcsec. The new weaker source Ace-3, which was not detected by \citet{Friedel2008}, coincides within 2\arcsec\ with the KL-W/MF-5 position \citep{Peng2012a,Favre2011a} and is close to the IR source IRc6. 

\item The overall distribution of acetone in BN/KL is more extended than the one found by \citet{Friedel2008} and is similar to that of the large N-bearing species (i.e., \ec). This is likely due to the higher sensitivity and different uv coverage of our PdBI data compared with the CARMA data of \citet{Friedel2008}.

\item Rotational temperatures of $159\pm14$ are determined toward Ace-1 and $145\pm16$ K toward Ace-2. The acetone column density toward these two peaks is about $2-4\times10^{16}$ \cmm\ with a relative abundance of $1-6\times10^{-8}$, within the range given by \citet{Friedel2005}. In addition, Acetone is a few times less abundant toward the hot core and Ace-3 peaks ($2-7\times10^{-9}$) compared with Ace-1 and Ace-2.

\end{enumerate}

Current chemical models require a formation of acetone on grain surfaces as for \de\ and they cannot explain the difference seen between the two species. Including reactions with N-bearing species either to form or to destroy acetone might be necessary to explain its different distribution \citep[e.g.,][]{Chen2011}. Additional laboratory and theoretical chemical studies are needed to progress in the understanding of the acetone abundance. High-resolution maps with more sensitive ALMA data will hopefully lead to the discovery of other species with the same spatial distribution, and confirm any association with shocks.


\begin{acknowledgements}
We thank the referee for the comments and helpful suggestions. T.-C. Peng acknowledges the support from the ALMA grant 2009-18 at Universit\'{e} de Bordeaux1/LAB. This paper makes use of the following ALMA data: ADS/JAO.ALMA\#2011.0.00009.SV. ALMA is a partnership of ESO (representing its member states), NSF (USA) and NINS (Japan), together with NRC (Canada) and NSC and ASIAA (Taiwan), in cooperation with the Republic of Chile. The Joint ALMA Observatory is operated by ESO, AUI/NRAO and NAOJ.
\end{acknowledgements}


\Online
\begin{appendix}
\section{Complementary figures and tables \label{app1}}


       \begin{figure*}
   \centering
   \includegraphics[angle=-90,width=0.45\textwidth]{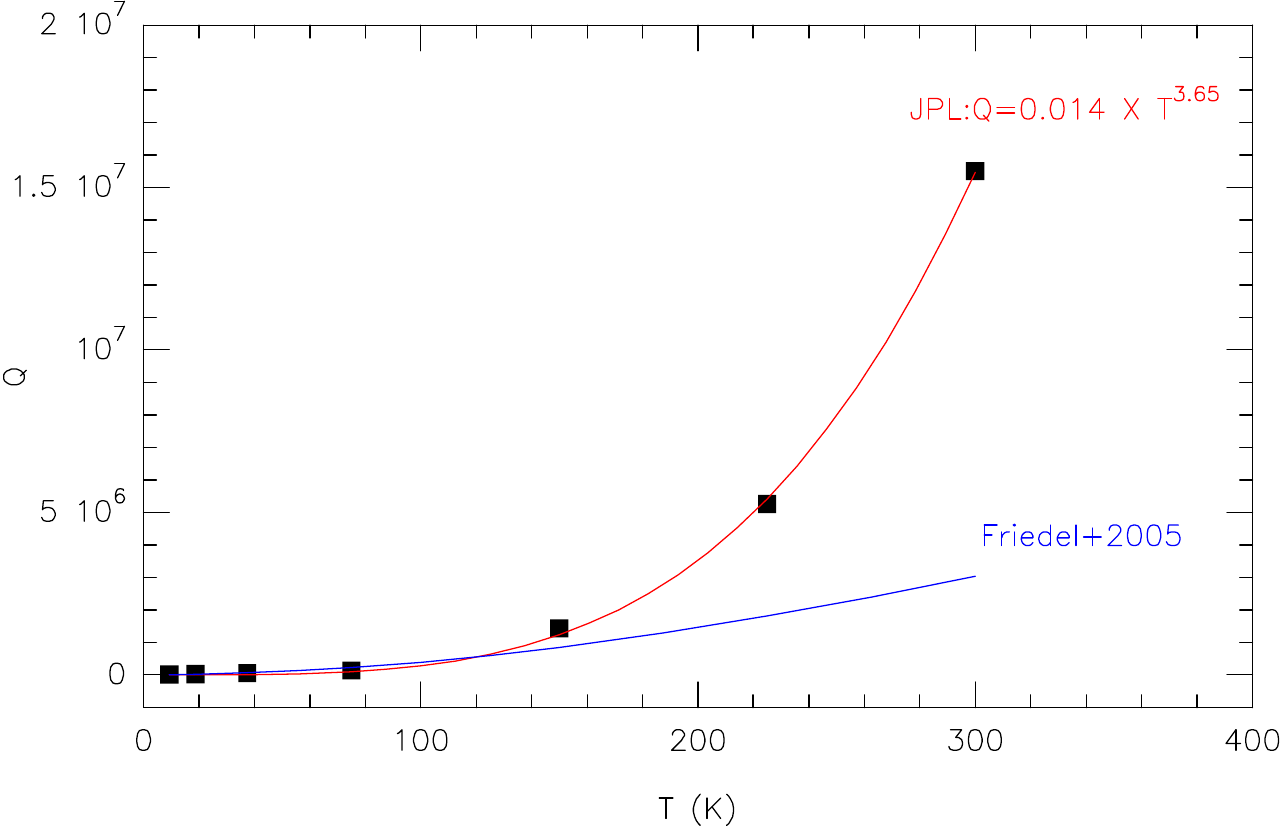}
   \caption{Comparison of the acetone partition functions taken from the JPL database (black squares) and the approximation (blue curve) used by \citet{Friedel2005}. The red curve is a fit to the JPL acetone partition functions.}
   \label{partition}
   \end{figure*}

\begin{table*}
\caption{\ace\ measurements toward Ace-1}             
\label{table-ace1}      
\centering                          
\begin{tabular}{llccccc}        
\hline\hline                 
Frequency  & Transition           &  $T_{\rm B}$  & $V_{\rm LSR}$ & $\Delta V$    & $\int T_{\rm B}dV$  & Comment\\    
(MHz)      & ($J_{k_{a},k_{c}}$)  &   (K)         & (km s$^{-1}$) & (km s$^{-1}$) & (K km s$^{-1}$)     & \\
       
\hline                        

101297.450 &$24_{13,12}-24_{12,13}$ EE & 0.85$\pm$0.05 & 7.31$\pm$1.85 & 5.55$\pm$1.85 &  4.97$\pm$0.83 & Partially blended$^1$  \\ 
\multicolumn{1}{|l}{101426.664} &      $9_{1,8}-8_{2,7}$ AE & \multirow{4}{*}{2.04$\pm$0.05} & \multirow{4}{*}{6.82$\pm$1.85} & \multirow{4}{*}{5.14$\pm$1.86} & \multirow{4}{*}{10.87$\pm$0.44} &     \\     
\multicolumn{1}{|l}{101426.759} &      $9_{1,8}-8_{2,7}$ EA & & & & &    \\
\multicolumn{1}{|l}{101427.041} &      $9_{2,8}-8_{1,7}$ AE & & & & &    \\
\multicolumn{1}{|l}{101427.130} &      $9_{2,8}-8_{1,7}$ EA &  & & & &    \\
\multicolumn{1}{l|}{101451.059} &      $9_{1,8}-8_{2,7}$ EE & \multirow{2}{*}{3.88$\pm$0.05} & \multirow{2}{*}{6.89$\pm$1.85} & \multirow{2}{*}{5.06$\pm$1.85} & \multirow{2}{*}{20.46$\pm$0.48} &   \\
\multicolumn{1}{l|}{101451.446} &      $9_{2,8}-8_{1,7}$ EE & & & & &     \\
\multicolumn{1}{|l}{101475.332} &      $9_{1,8}-8_{2,7}$ AA & \multirow{2}{*}{2.25$\pm$0.05} & \multirow{2}{*}{6.59$\pm$1.85} & \multirow{2}{*}{5.55$\pm$1.85} & \multirow{2}{*}{11.27$\pm$1.29} &  \multirow{2}{*}{Partially blended$^2$} \\
\multicolumn{1}{|l}{101475.734} &      $9_{2,8}-8_{1,7}$ AA & & & & &     \\

\multicolumn{1}{l|}{203336.247} &  $27_{4,23}-27_{3,24}$ EA & \multirow{4}{*}{14.30$\pm$1.19} & \multirow{4}{*}{8.27$\pm$0.92} & \multirow{4}{*}{3.48$\pm$0.94} & \multirow{4}{*}{51.41$\pm$4.57} & \multirow{4}{*}{Blended$^3$}\\
\multicolumn{1}{l|}{}	 	   &  $27_{5,23}-27_{4,24}$ EA & & & & &     \\
\multicolumn{1}{l|}{203336.291} &  $27_{4,23}-27_{4,24}$ AE & & & & &     \\
\multicolumn{1}{l|}{}	 	   &  $27_{5,23}-27_{3,24}$ AE & & & & &     \\ 
  
223512.409 &    $16_{8,9}-15_{7,8}$ EE & 4.37$\pm$0.37 & 7.29$\pm$0.84 & 2.96$\pm$0.86 &  14.20$\pm$1.33 &    Possibly blended$^4$ \\
223621.690 &    $16_{8,9}-15_{7,8}$ AA & 3.56$\pm$0.37 & 7.16$\pm$0.84 & 2.12$\pm$0.90 &   8.56$\pm$1.32 &    Blended$^5$ \\
\multicolumn{1}{|l}{223684.608} &  $17_{6,11}-16_{7,10}$ EA & \multirow{2}{*}{4.85$\pm$0.37} & \multirow{2}{*}{7.73$\pm$0.92} & \multirow{2}{*}{2.03$\pm$0.89} & \multirow{2}{*}{10.32$\pm$1.67} & \multirow{2}{*}{Blended$^6$}\\
\multicolumn{1}{|l}{223684.610} &  $17_{6,11}-16_{7,10}$ AE & & & & &     \\
\multicolumn{1}{l|}{223692.004} &  $17_{7,11}-16_{6,10}$ EA & \multirow{2}{*}{4.95$\pm$0.37} & \multirow{2}{*}{7.58$\pm$0.92} & \multirow{2}{*}{3.00$\pm$0.84} & \multirow{2}{*}{16.18$\pm$1.69} & \multirow{2}{*}{Partially blended$^7$}\\
\multicolumn{1}{l|}{223692.104} &  $17_{7,11}-16_{6,10}$ AE & & & & &     \\

223732.826 &    $12_{9,4}-11_{8,4}$ EE &  2.11$\pm$0.37 & 7.18$\pm$0.67 & 4.37$\pm$0.81 &   9.85$\pm$2.12 &    Blended$^8$\\
223767.585 &  $17_{6,11}-16_{7,10}$ EE & 11.61$\pm$0.37 & 7.53$\pm$0.84 & 2.92$\pm$0.85 &  35.27$\pm$1.79 &    Blended$^9$\\
223775.253 &  $17_{7,11}-16_{6,10}$ EE &  5.88$\pm$0.37 & 7.75$\pm$0.84 & 3.58$\pm$0.98 &  19.99$\pm$2.25 &    Possibly blended$^{10}$  \\
223850.417 &  $17_{6,11}-16_{7,10}$ AA &  3.84$\pm$0.37 & 7.13$\pm$0.84 & 3.30$\pm$0.84 &  14.13$\pm$1.32 &    Slightly blended$^{11}$ \\
223858.308 &  $17_{7,11}-16_{6,10}$ AA &  2.48$\pm$0.37 & 7.66$\pm$0.84 & 3.31$\pm$0.94 &   8.35$\pm$1.42 &    Possibly blended$^{12}$\\

\multicolumn{1}{|l}{225744.082} &  $19_{4,15}-18_{5,14}$ EE & \multirow{2}{*}{5.19$\pm$0.43} & \multirow{2}{*}{7.54$\pm$0.42} & \multirow{2}{*}{2.47$\pm$0.42} & \multirow{2}{*}{13.63$\pm$1.97} &     \\
\multicolumn{1}{|l}{}           &  $19_{5,15}-18_{4,14}$ EE & & & & &     \\

225811.979 &    $13_{9,5}-12_{8,4}$ EE &  1.81$\pm$0.34 & 8.46$\pm$0.42 & 2.99$\pm$0.47 &  5.76$\pm$0.64 & Possibly blended$^{13}$     \\
228979.750 &  $12_{10,2}-11_{9,2}$ EA  & ... & ... & ... & ... &  Blended$^{14}$\\

\multicolumn{1}{l|}{229033.736} &  $22_{1,21}-21_{2,20}$ AE & \multirow{4}{*}{4.47$\pm$0.64} & \multirow{4}{*}{7.38$\pm$0.45} & \multirow{4}{*}{2.80$\pm$0.41} & \multirow{4}{*}{13.33$\pm$1.26} &  \multirow{4}{*}{}\\
\multicolumn{1}{l|}{}           &  $22_{2,21}-21_{1,20}$ AE & & & & &     \\
\multicolumn{1}{l|}{229033.771} &  $22_{1,21}-21_{1,20}$ EA & & & & &     \\
\multicolumn{1}{l|}{}           &  $22_{2,21}-21_{2,20}$ EA & & & & &     \\

229041.826 &   $12_{10,3}-11_{9,3}$ EE &  3.01$\pm$0.62 & 8.46$\pm$0.53 & 3.30$\pm$0.41 &  10.55$\pm$1.38 &       \\

\multicolumn{1}{|l}{229055.797} &  $22_{1,21}-21_{2,20}$ EE & \multirow{4}{*}{6.92$\pm$0.90} & \multirow{4}{*}{7.47$\pm$0.44} & \multirow{4}{*}{3.07$\pm$0.52} & \multirow{4}{*}{22.58$\pm$2.68} &  \multirow{4}{*}{}\\
\multicolumn{1}{|l}{}           &  $22_{1,21}-21_{1,20}$ EE & & & & &     \\
\multicolumn{1}{|l}{}           &  $22_{2,21}-21_{2,20}$ EE & & & & &     \\
\multicolumn{1}{|l}{}           &  $22_{2,21}-21_{1,20}$ EE & & & & &     \\

229058.049 &    $14_{9,6}-13_{8,5}$ EE &  2.92$\pm$0.80 & 7.37$\pm$0.59 & 3.00$\pm$0.41 &  9.31$\pm$2.05 &     \\

\multicolumn{1}{l|}{229077.788} &  $22_{1,21}-21_{2,20}$ AA & \multirow{2}{*}{3.67$\pm$0.39} & \multirow{2}{*}{7.06$\pm$0.42} & \multirow{2}{*}{4.04$\pm$0.45} & \multirow{2}{*}{15.78$\pm$0.80} &  \multirow{2}{*}{}\\
\multicolumn{1}{l|}{}           &  $22_{2,21}-21_{1,20}$ AA & & & & &     \\

\hline                                   
\end{tabular}

\tablefoot{Acetone spectroscopic data are taken from the JPL database. Upper-state energy \Eup, effective line strength $S\mu^2$, and spin statistical weight $g_{\rm s}$ are also given.
\tablefoottext{1}{Partially blended with the NH$_2$CHO $18_{2,16}-18_{2,17}$ line at 101293.3 MHz.}
\tablefoottext{2}{Partially blended with the H$_2$CS $3_{1,3}-2_{1,2}$ line at 101477.8 MHz.}
\tablefoottext{3}{Blended with the NH$_2$CHO $10_{1,10}-9_{1,9}$ line at 203335.8 MHz.}
\tablefoottext{4}{Possibly blended with the CH$_3$CH$_2^{13}$CN $25_{4,22}-24_{4,21}$ line at 223512.8 MHz.}
\tablefoottext{5}{Partially blended with the \mf\ $37_{8,30}-37_{7,31}$ E line at 223624.5 MHz.}
\tablefoottext{6}{Blended with the U223686.7 line.}
\tablefoottext{7}{Partially blended with the U223694.8 line.}
\tablefoottext{8}{The emission peaks at about 7 \kms, which is likely be blended with the U223733.6 line.}
\tablefoottext{9}{Blended with the U223767.6 line.}
\tablefoottext{10}{Possibly blended with the $^{13}$CH$_3$CH$_2$CN $26_{2,25}-25_{2,24}$ line at 223773.5 MHz.}
\tablefoottext{11}{Slightly blended with the HCOOCH$_3$ $35_{7,29}-35_{6,30}$ A line at 223854.2 MHz.}
\tablefoottext{12}{Possibly blended with the $^{13}$CH$_3$CH$_2$CN $14_{2,12}-13_{1,13}$ line at 223859.3 MHz, similar to the 223775.3 MHz \ace\ line.}
\tablefoottext{13}{Possibly blended with the CH$_3^{13}$CH$_2$CN $27_{2,26}-27_{0,27}$ line at 225810.7 MHz.}
\tablefoottext{14}{Blended with the U228979.0 Line.}
}

\end{table*}

%
%

\begin{table*}
\caption{\ace\ measurements toward Ace-2}             
\label{table-ace2}      
\centering                          
\begin{tabular}{llccccc}        
\hline\hline                 
Frequency  & Transition           &  $T_{\rm B}$  & $V_{\rm LSR}$ & $\Delta V$    & $\int T_{\rm B}dV$  & Comment\\    
(MHz)      & ($J_{k_{a},k_{c}}$)  &   (K)         & (km s$^{-1}$) & (km s$^{-1}$) & (K km s$^{-1}$)     & \\
       
\hline                        

101297.450 &$24_{13,12}-24_{12,13}$ EE & 0.49$\pm$0.05 & 7.31$\pm$1.85 & 5.55$\pm$1.85 &  3.28$\pm$0.50 &  Partially blended$^1$ \\ 
\multicolumn{1}{|l}{101426.664} &      $9_{1,8}-8_{2,7}$ AE & \multirow{4}{*}{1.26$\pm$0.05} & \multirow{4}{*}{6.82$\pm$1.85} & \multirow{4}{*}{6.29$\pm$1.86} & \multirow{4}{*}{8.58$\pm$0.40} &     \\     
\multicolumn{1}{|l}{101426.759} &      $9_{1,8}-8_{2,7}$ EA & & & & &    \\
\multicolumn{1}{|l}{101427.041} &      $9_{2,8}-8_{1,7}$ AE & & & & &    \\
\multicolumn{1}{|l}{101427.130} &      $9_{2,8}-8_{1,7}$ EA &  & & & &    \\
\multicolumn{1}{l|}{101451.059} &      $9_{1,8}-8_{2,7}$ EE & \multirow{2}{*}{2.51$\pm$0.05} & \multirow{2}{*}{6.89$\pm$1.85} & \multirow{2}{*}{6.56$\pm$1.85} & \multirow{2}{*}{17.44$\pm$0.49} &   \\
\multicolumn{1}{l|}{101451.446} &      $9_{2,8}-8_{1,7}$ EE & & & & &     \\
\multicolumn{1}{|l}{101475.332} &      $9_{1,8}-8_{2,7}$ AA & \multirow{2}{*}{1.69$\pm$0.05} & \multirow{2}{*}{6.59$\pm$1.85} & \multirow{2}{*}{5.55$\pm$1.85} & \multirow{2}{*}{8.78$\pm$1.49} &  \multirow{2}{*}{Partially blended$^2$} \\
\multicolumn{1}{|l}{101475.734} &      $9_{2,8}-8_{1,7}$ AA & & & & &     \\

\multicolumn{1}{l|}{203336.247} &  $27_{4,23}-27_{3,24}$ EA & \multirow{4}{*}{1.78$\pm$1.19} & \multirow{4}{*}{8.27$\pm$0.92} & \multirow{4}{*}{4.93$\pm$0.94} & \multirow{4}{*}{9.35$\pm$5.74} & \multirow{4}{*}{Blended$^3$}\\
\multicolumn{1}{l|}{}	 	   &  $27_{5,23}-27_{4,24}$ EA & & & & &     \\
\multicolumn{1}{l|}{203336.291} &  $27_{4,23}-27_{4,24}$ AE & & & & &     \\
\multicolumn{1}{l|}{}	 	   &  $27_{5,23}-27_{3,24}$ AE & & & & &     \\ 
  
223512.409 &    $16_{8,9}-15_{7,8}$ EE & $<2.5$ & 6.50$\pm$0.84 & ... &  $<16.0$ &    Possibly blended$^4$\\
223621.690 &    $16_{8,9}-15_{7,8}$ AA & 1.88$\pm$0.37 & 7.16$\pm$0.84 & 2.36$\pm$0.86 &  4.91$\pm$0.86 &  Partially blended$^5$   \\

\multicolumn{1}{|l}{223684.608} &  $17_{6,11}-16_{7,10}$ EA & \multirow{2}{*}{$<3.1$} & \multirow{2}{*}{...} & \multirow{2}{*}{...} & \multirow{2}{*}{$<14.4$} &  \multirow{2}{*}{Blended$^6$}\\
\multicolumn{1}{|l}{223684.610} &  $17_{6,11}-16_{7,10}$ AE & & & & &     \\

\multicolumn{1}{l|}{223692.004} &  $17_{7,11}-16_{6,10}$ EA & \multirow{2}{*}{1.99$\pm$0.37} & \multirow{2}{*}{8.42$\pm$0.84} & \multirow{2}{*}{...} & \multirow{2}{*}{$<6.4$} &  \multirow{2}{*}{Partially Blended$^7$} \\
\multicolumn{1}{l|}{223692.104} &  $17_{7,11}-16_{6,10}$ AE & & & & &     \\

223732.826 &  $12_{9,4}-11_{8,4}$ EE   & $<0.4$        & ...           & ...           & $<1.2$          &     Blended$^8$\\
223767.585 &  $17_{6,11}-16_{7,10}$ EE & 3.50$\pm$0.37 & 8.36$\pm$0.84 & 5.10$\pm$0.91 &  18.94$\pm$2.14 &     Blended$^9$\\
223775.253 &  $17_{7,11}-16_{6,10}$ EE & 4.34$\pm$0.37 & 7.75$\pm$0.84 & 4.54$\pm$0.89 &  21.87$\pm$1.98 &     Possibly blended$^{10}$\\
223850.417 &  $17_{6,11}-16_{7,10}$ AA & 1.58$\pm$0.37 & 7.13$\pm$0.84 & 4.39$\pm$0.94 &   7.85$\pm$1.63 &     Blended$^{11}$\\
223858.308 &  $17_{7,11}-16_{6,10}$ AA & 1.58$\pm$0.37 & 7.66$\pm$0.84 & 3.67$\pm$0.87 &   6.06$\pm$1.39 &     Possibly blended$^{12}$\\

\multicolumn{1}{|l}{225744.082} &  $19_{4,15}-18_{5,14}$ EE & \multirow{2}{*}{2.04$\pm$0.46} & \multirow{2}{*}{7.35$\pm$0.43} & \multirow{2}{*}{3.71$\pm$0.48} & \multirow{2}{*}{8.08$\pm$2.99} &     \\
\multicolumn{1}{|l}{}           &  $19_{5,15}-18_{4,14}$ EE & & & & &     \\

225811.979 &    $13_{9,5}-12_{8,4}$ EE &  0.82$\pm$0.29 & 9.13$\pm$0.45 & 3.00$\pm$0.41 &  2.63$\pm$0.59 &   Possibly blended$^{13}$   \\

228979.750 &  $12_{10,2}-11_{9,2}$ EA  & ... & ... & ... & ... &  Blended$^{14}$\\

\multicolumn{1}{l|}{229033.736} &  $22_{1,21}-21_{2,20}$ AE & \multirow{4}{*}{$<1.2$} & \multirow{4}{*}{...} & \multirow{4}{*}{...} & \multirow{4}{*}{$<3.6$} &  \multirow{4}{*}{}\\
\multicolumn{1}{l|}{}           &  $22_{2,21}-21_{1,20}$ AE & & & & &     \\
\multicolumn{1}{l|}{229033.771} &  $22_{1,21}-21_{1,20}$ EA & & & & &     \\
\multicolumn{1}{l|}{}           &  $22_{2,21}-21_{2,20}$ EA & & & & &     \\

229041.826 &   $12_{10,3}-11_{9,3}$ EE & $<0.9$ & ... & ... & $<2.7$ &     \\

\multicolumn{1}{|l}{229055.797} &  $22_{1,21}-21_{2,20}$ EE & \multirow{4}{*}{2.70$\pm$0.47} & \multirow{4}{*}{8.01$\pm$0.45} & \multirow{4}{*}{3.00$\pm$0.41} & \multirow{4}{*}{8.62$\pm$0.78} &  \multirow{4}{*}{}\\
\multicolumn{1}{|l}{}           &  $22_{1,21}-21_{1,20}$ EE & & & & &     \\
\multicolumn{1}{|l}{}           &  $22_{2,21}-21_{2,20}$ EE & & & & &     \\
\multicolumn{1}{|l}{}           &  $22_{2,21}-21_{1,20}$ EE & & & & &     \\

229058.049 &    $14_{9,6}-13_{8,5}$ EE &  2.66$\pm$0.46 & 7.72$\pm$0.45 & 3.00$\pm$0.41 &  8.49$\pm$0.75 &     \\

\multicolumn{1}{l|}{229077.788} &  $22_{1,21}-21_{2,20}$ AA & \multirow{2}{*}{1.58$\pm$0.16} & \multirow{2}{*}{7.50$\pm$0.42} & \multirow{2}{*}{5.18$\pm$0.48} & \multirow{2}{*}{8.69$\pm$0.65} &  \multirow{2}{*}{}\\
\multicolumn{1}{l|}{}           &  $22_{2,21}-21_{1,20}$ AA & & & & &     \\

\hline                                   
\end{tabular}
\tablefoot{Acetone spectroscopic data are taken from the JPL database. Upper-state energy \Eup, effective line strength $S\mu^2$, and spin statistical weight $g_{\rm s}$ are also given.
\tablefoottext{1}{Partially blended with the NH$_2$CHO $18_{2,16}-18_{2,17}$ line at 101293.3 MHz.}
\tablefoottext{2}{Partially blended with the H$_2$CS $3_{1,3}-2_{1,2}$ line at 101477.8 MHz.}
\tablefoottext{3}{Blended with the NH$_2$CHO $10_{1,10}-9_{1,9}$ line at 203335.8 MHz.}
\tablefoottext{4}{Possibly blended with the CH$_3$CH$_2^{13}$CN $25_{4,22}-24_{4,21}$ line at 223512.8 MHz.}
\tablefoottext{5}{Partially blended with the \mf\ $37_{8,30}-37_{7,31}$ E line at 223624.5 MHz.}
\tablefoottext{6}{Blended with the strong U223680.0 line.}
\tablefoottext{7}{Partially blended with the U223694.8 line.}
\tablefoottext{8}{Blended with the U223733.6 line.}
\tablefoottext{9}{Blended with the U223767.6 line.}
\tablefoottext{10}{Possibly blended with the $^{13}$CH$_3$CH$_2$CN $26_{2,25}-25_{2,24}$ line at 223773.5 MHz.}
\tablefoottext{11}{Blended with the HCOOCH$_3$ $35_{7,29}-35_{6,30}$ A line at 223854.2 MHz.}
\tablefoottext{12}{Possibly blended with the $^{13}$CH$_3$CH$_2$CN $14_{2,12}-13_{1,13}$ line at 223859.3 MHz, similar to the 223775.3 MHz \ace\ line.}
\tablefoottext{13}{Possibly blended with the CH$_3^{13}$CH$_2$CN $27_{2,26}-27_{0,27}$ line at 225810.7 MHz.}
\tablefoottext{14}{Blended with the U228979.0 Line.}
}

\end{table*}

%
%

\begin{table*}
\caption{\ace\ measurements toward Ace-3}             
\label{table-ace3}      
\centering                          
\begin{tabular}{llccccc}        
\hline\hline                 
Frequency  & Transition           &  $T_{\rm B}$  & $V_{\rm LSR}$ & $\Delta V$    & $\int T_{\rm B}dV$  & Comment\\    
(MHz)      & ($J_{k_{a},k_{c}}$)  &   (K)         & (km s$^{-1}$) & (km s$^{-1}$) & (K km s$^{-1}$)     & \\
       
\hline                        

101297.450 &$24_{13,12}-24_{12,13}$ EE & 0.20$\pm$0.05 & 7.31$\pm$1.85 & 5.55$\pm$1.85 &  1.43$\pm$0.63 &   \\ 
\multicolumn{1}{|l}{101426.664} &      $9_{1,8}-8_{2,7}$ AE & \multirow{4}{*}{0.72$\pm$0.05} & \multirow{4}{*}{6.82$\pm$1.85} & \multirow{4}{*}{7.10$\pm$1.86} & \multirow{4}{*}{5.78$\pm$0.46} &     \\     
\multicolumn{1}{|l}{101426.759} &      $9_{1,8}-8_{2,7}$ EA & & & & &    \\
\multicolumn{1}{|l}{101427.041} &      $9_{2,8}-8_{1,7}$ AE & & & & &    \\
\multicolumn{1}{|l}{101427.130} &      $9_{2,8}-8_{1,7}$ EA &  & & & &    \\
\multicolumn{1}{l|}{101451.059} &      $9_{1,8}-8_{2,7}$ EE & \multirow{2}{*}{1.47$\pm$0.05} & \multirow{2}{*}{6.89$\pm$1.85} & \multirow{2}{*}{5.94$\pm$1.85} & \multirow{2}{*}{9.79$\pm$0.38} &   \\
\multicolumn{1}{l|}{101451.446} &      $9_{2,8}-8_{1,7}$ EE & & & & &     \\
\multicolumn{1}{|l}{101475.332} &      $9_{1,8}-8_{2,7}$ AA & \multirow{2}{*}{1.72$\pm$0.05} & \multirow{2}{*}{6.59$\pm$1.85} & \multirow{2}{*}{5.55$\pm$1.85} & \multirow{2}{*}{9.63$\pm$0.48} &  \multirow{2}{*}{Partially blended$^1$} \\
\multicolumn{1}{|l}{101475.734} &      $9_{2,8}-8_{1,7}$ AA & & & & &     \\

\multicolumn{1}{l|}{203336.247} &  $27_{4,23}-27_{3,24}$ EA & \multirow{4}{*}{$<$1.2} & \multirow{4}{*}{...} & \multirow{4}{*}{...} & \multirow{4}{*}{$<$2.2} &     \\
\multicolumn{1}{l|}{}	 	   &  $27_{5,23}-27_{4,24}$ EA & & & & &     \\
\multicolumn{1}{l|}{203336.291} &  $27_{4,23}-27_{4,24}$ AE & & & & &     \\
\multicolumn{1}{l|}{}	 	   &  $27_{5,23}-27_{3,24}$ AE & & & & &     \\ 
  
223512.409 &    $16_{8,9}-15_{7,8}$ EE & $<0.4$ & ... & ... & $<1.2$ &    \\
223621.690 &    $16_{8,9}-15_{7,8}$ AA & $<0.4$ & ... & ... & $<1.2$ &    \\

\multicolumn{1}{|l}{223684.608} &  $17_{6,11}-16_{7,10}$ EA & \multirow{2}{*}{$<0.4$} & \multirow{2}{*}{...} & \multirow{2}{*}{...} & \multirow{2}{*}{$<1.2$} & \multirow{2}{*}{blended$^2$} \\
\multicolumn{1}{|l}{223684.610} &  $17_{6,11}-16_{7,10}$ AE & & & & &     \\

\multicolumn{1}{l|}{223692.004} &  $17_{7,11}-16_{6,10}$ EA & \multirow{2}{*}{$<0.4$} & \multirow{2}{*}{...} & \multirow{2}{*}{...} & \multirow{2}{*}{$<1.2$} &   \\
\multicolumn{1}{l|}{223692.104} &  $17_{7,11}-16_{6,10}$ AE & & & & &     \\

223732.826 &    $12_{9,4}-11_{8,4}$ EE &  $<0.5$ & ... & ... & $<1.3$ &    Possibly blended$^3$\\
223767.585 &  $17_{6,11}-16_{7,10}$ EE & 1.37$\pm$0.35 &  5.02$\pm$0.84 & 1.73$\pm$0.86 &  2.66$\pm$0.62 &  first component\\
           &                           & 1.19$\pm$0.35 & 10.00$\pm$0.84 & 1.65$\pm$0.86 &  2.23$\pm$0.59 &  second component   \\
223775.253 &  $17_{7,11}-16_{6,10}$ EE & 1.29$\pm$0.35 &  6.07$\pm$0.84 & 1.51$\pm$0.86 &  2.08$\pm$0.55 &  first component   \\
           &                           & 0.83$\pm$0.35 & 10.26$\pm$0.84 & 2.86$\pm$0.93 &  2.25$\pm$1.00 &  second component$^4$    \\
223850.417 &  $17_{6,11}-16_{7,10}$ AA & $<0.4$ & ... & ... & $<1.2$ &     \\
223858.308 &  $17_{7,11}-16_{6,10}$ AA & $<0.4$ & ... & ... & $<1.2$ &     \\

\multicolumn{1}{|l}{225744.082} &  $19_{4,15}-18_{5,14}$ EE & \multirow{2}{*}{$<1.2$} & \multirow{2}{*}{...} & \multirow{2}{*}{...} & \multirow{2}{*}{$<3.6$} &     \\
\multicolumn{1}{|l}{}           &  $19_{5,15}-18_{4,14}$ EE & & & & &     \\

225811.979 &    $13_{9,5}-12_{8,4}$ EE &  $<0.2$ & ... & ... &  $<0.6$ &      \\

228979.750 &  $12_{10,2}-11_{9,2}$ EA  & ... & ... & ... & ... &  Blended$^5$\\

\multicolumn{1}{l|}{229033.736} &  $22_{1,21}-21_{2,20}$ AE & \multirow{4}{*}{1.04$\pm$0.33} & \multirow{4}{*}{7.18$\pm$0.45} & \multirow{4}{*}{3.20$\pm$0.41} & \multirow{4}{*}{3.54$\pm$0.53} &  \multirow{4}{*}{}\\
\multicolumn{1}{l|}{}           &  $22_{2,21}-21_{1,20}$ AE & & & & &     \\
\multicolumn{1}{l|}{229033.771} &  $22_{1,21}-21_{1,20}$ EA & & & & &     \\
\multicolumn{1}{l|}{}           &  $22_{2,21}-21_{2,20}$ EA & & & & &     \\
229041.826 &   $12_{10,3}-11_{9,3}$ EE & $<0.5$ & ... & ... & $<1.5$ &     \\

\multicolumn{1}{|l}{229055.797} &  $22_{1,21}-21_{2,20}$ EE & \multirow{4}{*}{$<1.3$} & \multirow{4}{*}{...} & \multirow{4}{*}{...} & \multirow{4}{*}{$<3.9$} &  \multirow{4}{*}{}\\
\multicolumn{1}{|l}{}           &  $22_{1,21}-21_{1,20}$ EE & & & & &     \\
\multicolumn{1}{|l}{}           &  $22_{2,21}-21_{2,20}$ EE & & & & &     \\
\multicolumn{1}{|l}{}           &  $22_{2,21}-21_{1,20}$ EE & & & & &     \\

229058.049 &    $14_{9,6}-13_{8,5}$ EE & $<0.9$ & ... & ... & $<2.7$ &     \\

\multicolumn{1}{l|}{229077.788} &  $22_{1,21}-21_{2,20}$ AA & \multirow{2}{*}{1.24$\pm$0.39} & \multirow{2}{*}{7.75$\pm$0.56} & \multirow{2}{*}{3.00$\pm$0.41} & \multirow{2}{*}{3.94$\pm$0.59} &  \multirow{2}{*}{}\\
\multicolumn{1}{l|}{}           &  $22_{2,21}-21_{1,20}$ AA & & & & &     \\

\hline                                   
\end{tabular}

\tablefoot{Acetone spectroscopic data are taken from the JPL database. Upper-state energy \Eup, effective line strength $S\mu^2$, and spin statistical weight $g_{\rm s}$ are also given. The two velocity components in Ace-3 may be due to an incomplete uv coverage so were measured separately.  
\tablefoottext{1}{Partially blended with the H$_2$CS $3_{1,3}-2_{1,2}$ line at 101477.8 MHz.}
\tablefoottext{2}{Blended with the U223686.7 line.}
\tablefoottext{3}{Possibly blended with the U223733.6 line.}
\tablefoottext{4}{Possible blended with the $^{13}$CH$_3$CH$_2$CN $26_{2,25}-25_{2,24}$ line at 223773.5 MHz.}
\tablefoottext{5}{Blended with the U228979.0 Line.}
}
\end{table*}

%
%

\begin{table*}
\caption{\ace\ measurements toward HC}             
\label{table-hc}      
\centering                          
\begin{tabular}{llccccc}        
\hline\hline                 
Frequency  & Transition           &  $T_{\rm B}$  & $V_{\rm LSR}$ & $\Delta V$    & $\int T_{\rm B}dV$  & Comment\\    
(MHz)      & ($J_{k_{a},k_{c}}$)  &   (K)         & (km s$^{-1}$) & (km s$^{-1}$) & (K km s$^{-1}$)     & \\
       
\hline                        

101297.450 &$24_{13,12}-24_{12,13}$ EE & 0.42$\pm$0.05 & 7.31$\pm$1.85 & 5.55$\pm$1.85 &  1.43$\pm$0.63 &  Partially blended$^1$ \\ 
\multicolumn{1}{|l}{101426.664} &      $9_{1,8}-8_{2,7}$ AE & \multirow{4}{*}{0.78$\pm$0.05} & \multirow{4}{*}{6.82$\pm$1.85} & \multirow{4}{*}{6.21$\pm$1.87} & \multirow{4}{*}{2.62$\pm$0.53} &     \\     
\multicolumn{1}{|l}{101426.759} &      $9_{1,8}-8_{2,7}$ EA & & & & &    \\
\multicolumn{1}{|l}{101427.041} &      $9_{2,8}-8_{1,7}$ AE & & & & &    \\
\multicolumn{1}{|l}{101427.130} &      $9_{2,8}-8_{1,7}$ EA &  & & & &    \\
\multicolumn{1}{l|}{101451.059} &      $9_{1,8}-8_{2,7}$ EE & \multirow{2}{*}{1.76$\pm$0.05} & \multirow{2}{*}{6.89$\pm$1.85} & \multirow{2}{*}{6.18$\pm$1.85} & \multirow{2}{*}{11.93$\pm$0.42} &   \\
\multicolumn{1}{l|}{101451.446} &      $9_{2,8}-8_{1,7}$ EE & & & & &     \\
\multicolumn{1}{|l}{101475.332} &      $9_{1,8}-8_{2,7}$ AA & \multirow{2}{*}{1.15$\pm$0.05} & \multirow{2}{*}{6.59$\pm$1.85} & \multirow{2}{*}{5.55$\pm$1.85} & \multirow{2}{*}{5.07$\pm$0.86} &  \multirow{2}{*}{Partially blended$^2$} \\
\multicolumn{1}{|l}{101475.734} &      $9_{2,8}-8_{1,7}$ AA & & & & &     \\

\multicolumn{1}{l|}{203336.247} &  $27_{4,23}-27_{3,24}$ EA & \multirow{4}{*}{6.70$\pm$1.19} & \multirow{4}{*}{7.35$\pm$0.92} & \multirow{4}{*}{4.98$\pm$0.94} & \multirow{4}{*}{44.73$\pm$6.04} & \multirow{4}{*}{Blended$^3$} \\
\multicolumn{1}{l|}{}	 	   &  $27_{5,23}-27_{4,24}$ EA & & & & &     \\
\multicolumn{1}{l|}{203336.291} &  $27_{4,23}-27_{4,24}$ AE & & & & &     \\
\multicolumn{1}{l|}{}	 	   &  $27_{5,23}-27_{3,24}$ AE & & & & &     \\ 
  
223512.409 &    $16_{8,9}-15_{7,8}$ EE & 2.03$\pm$0.35 &  4.77$\pm$0.84 & ...           &  $<12.7$       &   Possibly  blended$^4$ \\
223621.690 &    $16_{8,9}-15_{7,8}$ AA & 2.37$\pm$0.37 &  6.32$\pm$0.84 & 4.45$\pm$0.92 & 10.53$\pm$1.70 &     \\

\multicolumn{1}{|l}{223684.608} &  $17_{6,11}-16_{7,10}$ EA & \multirow{2}{*}{$<5.4$} & \multirow{2}{*}{...} & \multirow{2}{*}{...} & \multirow{2}{*}{$<16.2$} & \multirow{2}{*}{blended$^6$}\\
\multicolumn{1}{|l}{223684.610} &  $17_{6,11}-16_{7,10}$ AE & & & & &     \\

\multicolumn{1}{l|}{223692.004} &  $17_{7,11}-16_{6,10}$ EA & \multirow{2}{*}{2.07$\pm$0.37} & \multirow{2}{*}{5.07$\pm$1.85} & \multirow{2}{*}{...} & \multirow{2}{*}{$<14.7$} &     \\
\multicolumn{1}{l|}{223692.104} &  $17_{7,11}-16_{6,10}$ AE & & & & &     \\

223732.826 &    $12_{9,4}-11_{8,4}$ EE & $<1.1$ & ... & ... & $<3.3$ &     blended$^7$ \\
223767.585 &  $17_{6,11}-16_{7,10}$ EE & 2.77$\pm$0.37 &  5.02$\pm$0.84 & 5.38$\pm$0.98 & 15.43$\pm$2.18 &      \\
223775.253 &  $17_{7,11}-16_{6,10}$ EE & 2.60$\pm$0.37 &  6.91$\pm$0.84 & 7.31$\pm$1.04 & 18.56$\pm$2.77 &  Possibly blended$^8$   \\
223850.417 &  $17_{6,11}-16_{7,10}$ AA & 0.82$\pm$0.37 &  6.30$\pm$0.84 & 6.40$\pm$1.08 &  5.18$\pm$2.19 &     \\
223858.308 &  $17_{7,11}-16_{6,10}$ AA & 0.81$\pm$0.37 &  5.99$\pm$0.84 & 4.95$\pm$0.97 &  4.04$\pm$1.70 &  Possibly blended$^9$   \\

\multicolumn{1}{|l}{225744.082} &  $19_{4,15}-18_{5,14}$ EE & \multirow{2}{*}{$<2.0$} & \multirow{2}{*}{...} & \multirow{2}{*}{...} & \multirow{2}{*}{$<10.0$} &     \\
\multicolumn{1}{|l}{}           &  $19_{5,15}-18_{4,14}$ EE & & & & &     \\

225811.979 &    $13_{9,5}-12_{8,4}$ EE &  $<0.4$ & ... & ... &  $<1.2$ &   Possibly blended$^{10}$   \\
228979.750 &  $12_{10,2}-11_{9,2}$ EA  & ... & ... & ... & ...  &  Blended$^{11}$\\
\multicolumn{1}{l|}{229033.736} &  $22_{1,21}-21_{2,20}$ AE & \multirow{4}{*}{2.72$\pm$0.63} & \multirow{4}{*}{7.74$\pm$0.41} & \multirow{4}{*}{3.20$\pm$0.41} & \multirow{4}{*}{9.27$\pm$1.38} &  \multirow{4}{*}{}\\
\multicolumn{1}{l|}{}           &  $22_{2,21}-21_{1,20}$ AE & & & & &     \\
\multicolumn{1}{l|}{229033.771} &  $22_{1,21}-21_{1,20}$ EA & & & & &     \\
\multicolumn{1}{l|}{}           &  $22_{2,21}-21_{2,20}$ EA & & & & &     \\

229041.826 &   $12_{10,3}-11_{9,3}$ EE & 1.98$\pm$0.63 &  8.13$\pm$0.69 & 3.20$\pm$0.41 & 6.75$\pm$1.38 &     \\

\multicolumn{1}{|l}{229055.797} &  $22_{1,21}-21_{2,20}$ EE & \multirow{4}{*}{4.50$\pm$0.52} & \multirow{4}{*}{7.52$\pm$0.43} & \multirow{4}{*}{3.00$\pm$0.41} & \multirow{4}{*}{14.37$\pm$0.93} &  \multirow{4}{*}{}\\
\multicolumn{1}{|l}{}           &  $22_{1,21}-21_{1,20}$ EE & & & & &     \\
\multicolumn{1}{|l}{}           &  $22_{2,21}-21_{2,20}$ EE & & & & &     \\
\multicolumn{1}{|l}{}           &  $22_{2,21}-21_{1,20}$ EE & & & & &     \\

229058.049 &    $14_{9,6}-13_{8,5}$ EE & 3.03$\pm$0.52 &  7.25$\pm$0.46 & 3.00$\pm$0.41 &  9.67$\pm$0.92 &    \\

\multicolumn{1}{l|}{229077.788} &  $22_{1,21}-21_{2,20}$ AA & \multirow{2}{*}{2.19$\pm$0.47} & \multirow{2}{*}{6.90$\pm$0.42} & \multirow{2}{*}{4.38$\pm$0.45} & \multirow{2}{*}{10.21$\pm$1.14} &  \multirow{2}{*}{}\\
\multicolumn{1}{l|}{}           &  $22_{2,21}-21_{1,20}$ AA & & & & &     \\

\hline                                   
\end{tabular}

\tablefoot{Acetone spectroscopic data are taken from the JPL database. Upper-state energy \Eup, effective line strength $S\mu^2$, and spin statistical weight $g_{\rm s}$ are also given.
\tablefoottext{1}{Partially blended with the NH$_2$CHO $18_{2,16}-18_{2,17}$ line at 101293.3 MHz.}
\tablefoottext{2}{Partially blended with the H$_2$CS $3_{1,3}-2_{1,2}$ line at 101477.8 MHz.}
\tablefoottext{3}{Blended with the NH$_2$CHO $10_{1,10}-9_{1,9}$ line at 203335.8 MHz.}
\tablefoottext{4}{Possibly blended with the CH$_3$CH$_2^{13}$CN $25_{4,22}-24_{4,21}$ line at 223512.8 MHz.}
\tablefoottext{5}{Partially blended with the \mf\ $37_{8,30}-37_{7,31}$ E line at 223624.5 MHz.}
\tablefoottext{6}{Blended with the strong U223680.0 line.}
\tablefoottext{7}{Blended with the U223733.6 line.}
\tablefoottext{8}{Possibly blended with the $^{13}$CH$_3$CH$_2$CN $26_{2,25}-25_{2,24}$ line at 223773.5 MHz.}
\tablefoottext{9}{Possibly blended with the $^{13}$CH$_3$CH$_2$CN $14_{2,12}-13_{1,13}$ line at 223859.3 MHz, similar to 223775.3 MHz \ace\ line.}
\tablefoottext{10}{Possibly blended with the CH$_3^{13}$CH$_2$CN $27_{2,26}-27_{0,27}$ line at 225810.7 MHz.}
\tablefoottext{11}{Blended with the U228979.0 Line.}
}

\end{table*}


      \begin{figure*}
   \centering
   \includegraphics[angle=-90,width=0.7\textwidth]{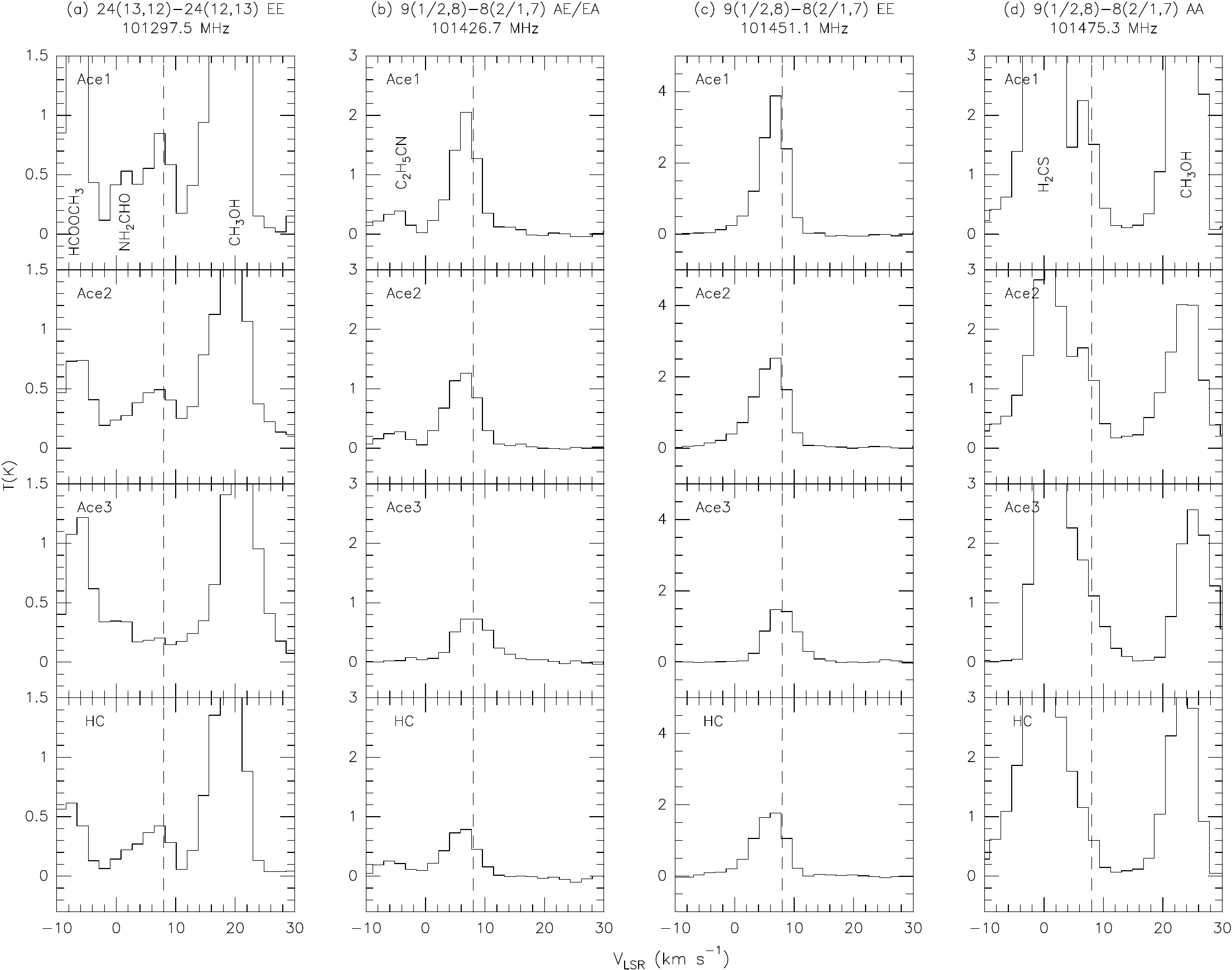}
   \caption{\ace\ spectra near 101 GHz toward four positions (three acetone emission peaks and HC) in the Orion BN/KL region. (b)-(c) Spectra correspond to the channel maps shown in Fig. \ref{Fig-ace-chmap-1} at 101427 MHz and 101451 MHz, respectively. Dashed lines indicate the \vlsr\ of 8 \kms\ and correspond to the \ace\ line shown at the top of each panel.}
   \label{Spectra-1}
   \end{figure*}

          \begin{figure*}
   \centering
   \includegraphics[angle=-90,width=0.8\textwidth]{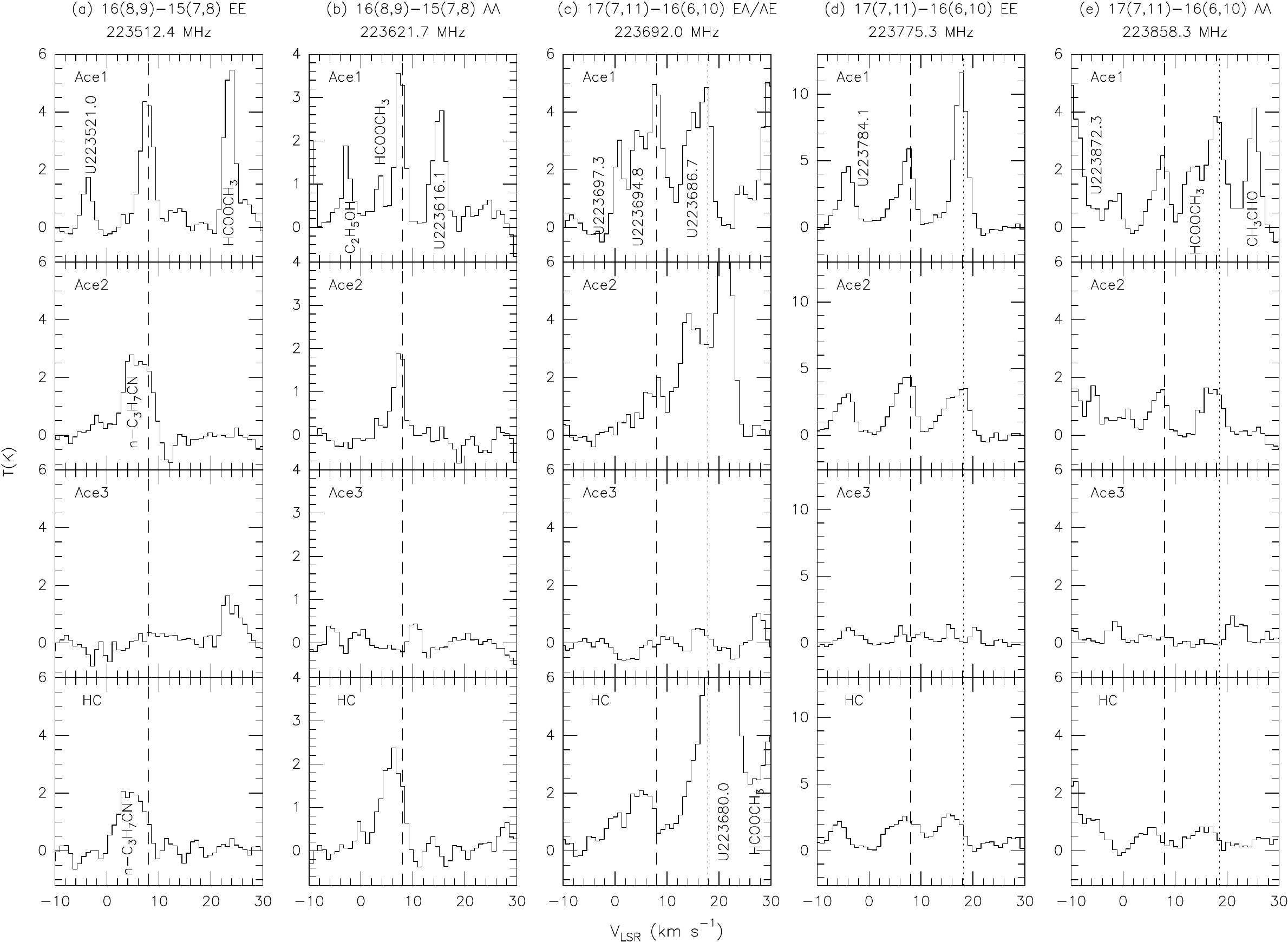}
   \caption{\ace\ spectra near 223 GHz toward four positions (three acetone emission peaks and HC) in the Orion BN/KL region. Dashed lines indicate the \vlsr\ of 8 \kms\ and correspond to the \ace\ line shown at the top of each panel. (c) The dotted line indicates the \ace\ $17_{6,11}-16_{7,10}$ EA/AE lines at 223684.6 MHz. (d) The dotted line indicates the \ace\ $17_{6,11}-16_{7,10}$ EE line at 223767.6 MHz. (e) The dotted line indicates the \ace\ $17_{6,11}-16_{7,10}$ AA line at 223850.4 MHz.}
   \label{Spectra-2}
   \end{figure*}

             \begin{figure*}
   \centering
   \includegraphics[angle=-90,width=0.8\textwidth]{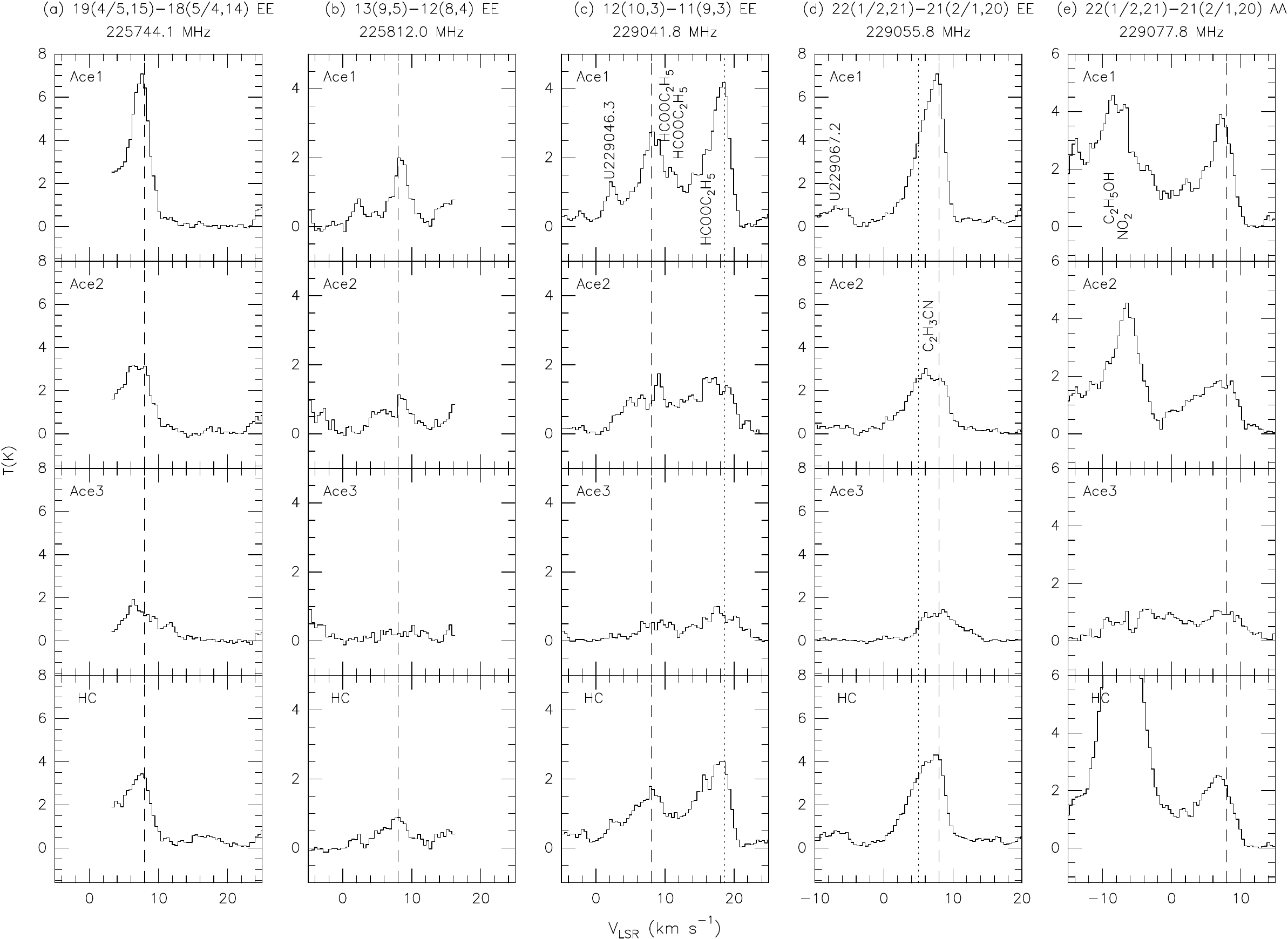}
   \caption{\ace\ spectra near 225 and 229 GHz toward four positions (three acetone emission peaks and HC) in the Orion BN/KL region. Dashed lines indicate the \vlsr\ of 8 \kms\ and correspond to the \ace\ line shown at the top of each panel. (c) The dotted line indicates the \ace\ $22_{1,21}-21_{2,20}$ AE/EA and $22_{2,21}-21_{1,20}$ AE/EA lines at 229033.8 MHz. (d) The dotted line indicates the \ace\ $14_{9,6}-13_{8,5}$ EE line at 229058.0 MHz.}
   \label{Spectra-3}
   \end{figure*}

\clearpage

\section{Comparison with the ALMA science verification data \label{app2}}

\begin{table*}
\caption{Comparison of the ALMA-SV and PdBI observational parameters at 223.7 GHz}             
\label{table-alma}      
\centering                          
\tabcolsep=0.11cm
\begin{tabular}{lcccccccc}        
\hline\hline                 
Data  & Antenna number & Integration time\tablefootmark{a}    & Flux conversion  & RMS noise  & $\theta_{\rm HPBW}$\tablefootmark{b} & $\delta{\rm v}$\tablefootmark{c}  & \multicolumn{2}{c}{$\theta_{\rm syn}$\tablefootmark{d}} \\    
      &                &   (min)             & (1 Jy beam$^{-1}$)  & (\mjb)     & (\arcsec)                            & (\kms)                            & (\arcsec$\times$\arcsec)  & PA (\degr)              \\
\hline                        

PdBI     &  5   & $\sim150$  & 17.3 K  & 21.2 &  22.5  & 0.84  & $1.79\times0.79$  & 14 \\
ALMA-SV  &  12  & $\sim100$  &  9.2 K  &  9.6 &  28.2  & 0.65  & $1.92\times1.39$  & 166 \\

\hline                                   
\end{tabular}
\tablefoot{
\tablefoottext{a}{On-source integration time}
\tablefoottext{b}{Primary beam size}
\tablefoottext{c}{Channel separation}
\tablefoottext{d}{Synthesized beam size}
}
\end{table*}

\subsection*{Basic observational parameters}

The calibrated ALMA-SV data were retrieved from the ESO archive. The observations were carried out in 2012 January with 16 12-m antennas \citep[see e.g.,][for more observational details]{Zapata2012,Niederhofer2012}. The primary beam at 223 GHz is about 28\arcsec. Some observational parameters of the ALMA-SV and our PdBI data are listed in Table \ref{table-alma}. The ALMA-SV data were analyzed with the GILDAS package, and the acetone emission images were cleaned with the same Clark algorithm as for the PdBI data to minimize the effects contributed by different cleaning algorithms. The ALMA and PdBI channel maps of the acetone $17_{7,11}-16_{6,10}$ EE line are shown in Figure \ref{pdbi-alma-chmap}, and the spectrum comparison around 223 GHz toward different positions is shown in Figures \ref{pdbi-alma-spectra-1}-\ref{pdbi-alma-spectra-3}. The missing flux of the ALMA-SV data is estimated to be 10--20\% compared with the IRAM 30m single-dish data (J. Cernicharo, priv. comm.) around 223 GHz, less than our PdBI data (20-50\% missing flux).

\subsection*{Spatial comparison}

\begin{figure*}
   \centering
   \includegraphics[angle=-90,width=0.7\textwidth]{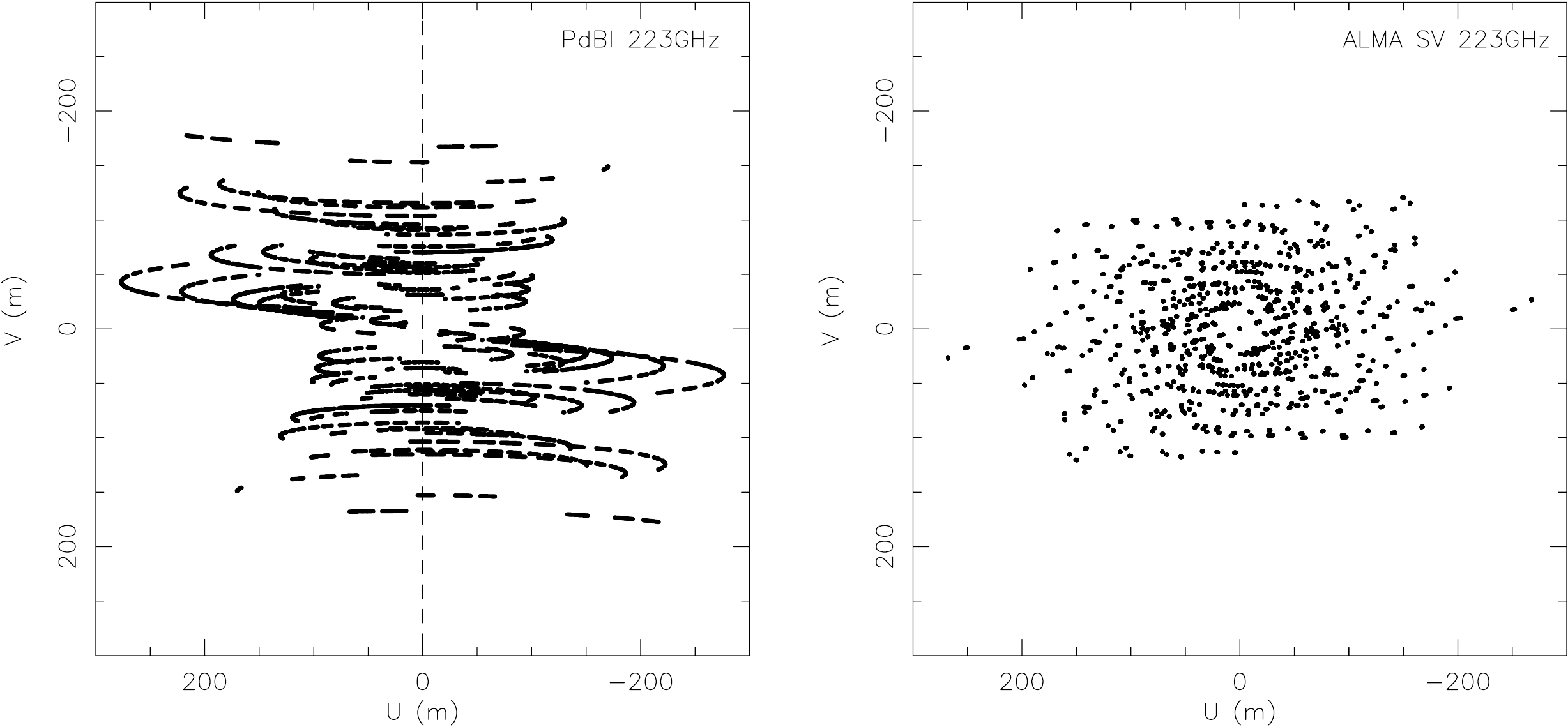}
   \caption{Comparison of the PdBI and ALMA-SV uv coverages.}
   \label{uvcover}
   \end{figure*}

\begin{figure*}
   \centering
   \includegraphics[angle=-90,width=0.8\textwidth]{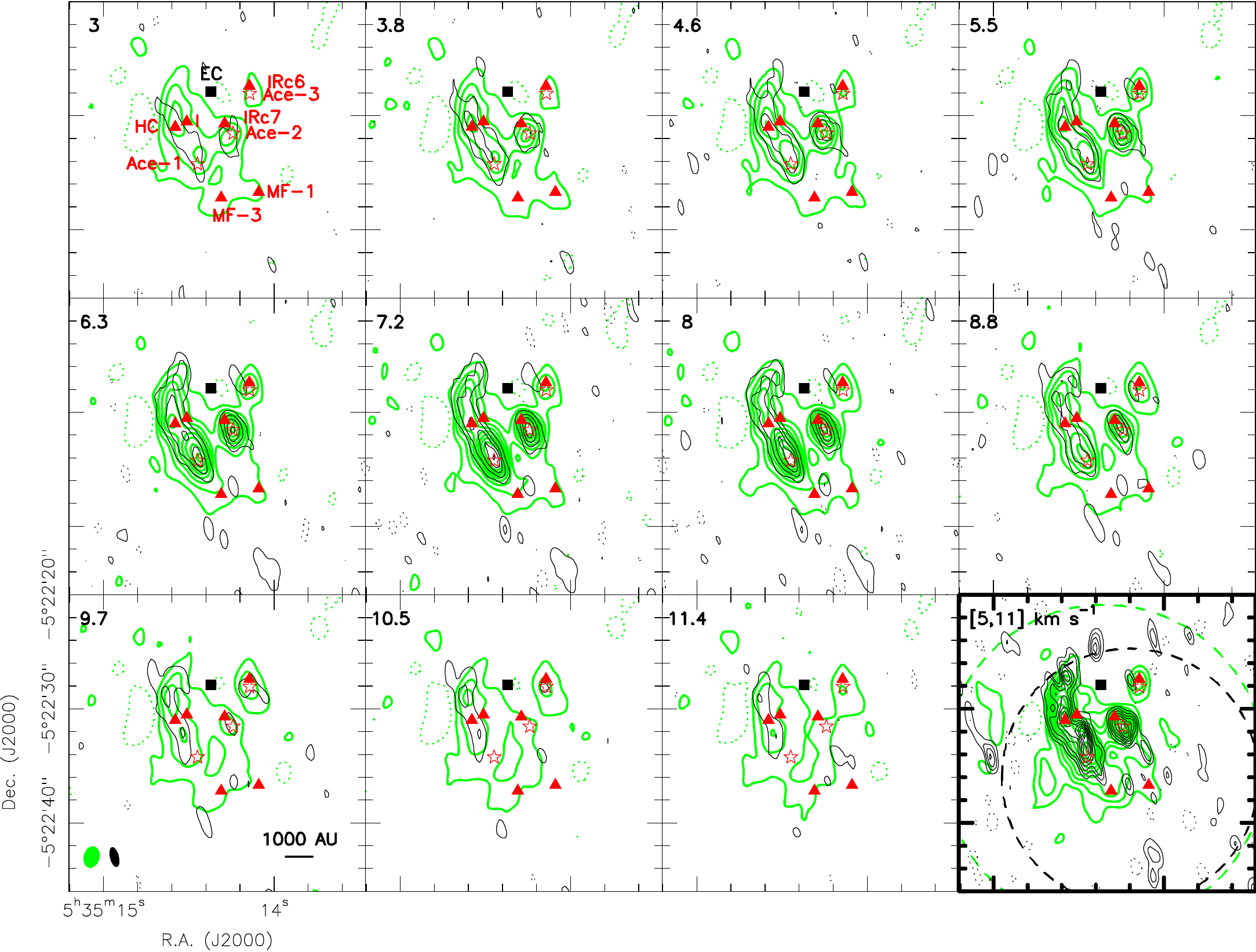}
   \caption{Comparison of the PdBI (black) and ALMA-SV (green) channel maps of the acetone $17_{7,11}-16_{6,10}$ EE line (\Eup=110.7 K) at 223775.3 MHz. Contours run from 0.6 to 8.1 K in steps of 0.9 K, and the dashed contours represent --0.6 K. The bottom-right panel shows the integrated intensity (from 5 to 11 \kms) in contours running from 15\% to 95\% in steps of 10\% of their peak temperatures, and the dashed contours represent --10\% of their peak temperatures. The primary beams of the ALMA-SV and PdBI data are indicated. The black square marks the center of explosion according to \citet{Zapata2009}. The positions of source BN, HC, IRc6/7, and source $I$ are marked as triangles. The positions of acetone emission peaks (Ace-1 to Ace-3) are marked as stars.}
   \label{pdbi-alma-chmap}
   \end{figure*}

Figure \ref{uvcover} shows the uv coverages of the PdBI and ALMA-SV data. They are comparable, but the ALMA-SV data show a denser coverage for shorter spacings. The PdBI data have longer baselines, which result in a higher angular resolution. However, due to the few uv tracks in the $-$50 to 50 m domain, the PdBI observations are likely to filter out larger structures in BN/KL. This is clearly seen in Figure \ref{pdbi-alma-chmap} in which most of the emission toward Ace-3 and in the southern part of BN/KL close to MF-1 and MF-3 is filtered out in the PdBI data. Likewise, it is important to note that the acetone emission associated with the arc-like structure in the north of HC is not resolved out by the PdBI.


\subsection*{Spectral comparison}
\begin{figure*}
   \centering
   \includegraphics[angle=0,width=0.8\textwidth]{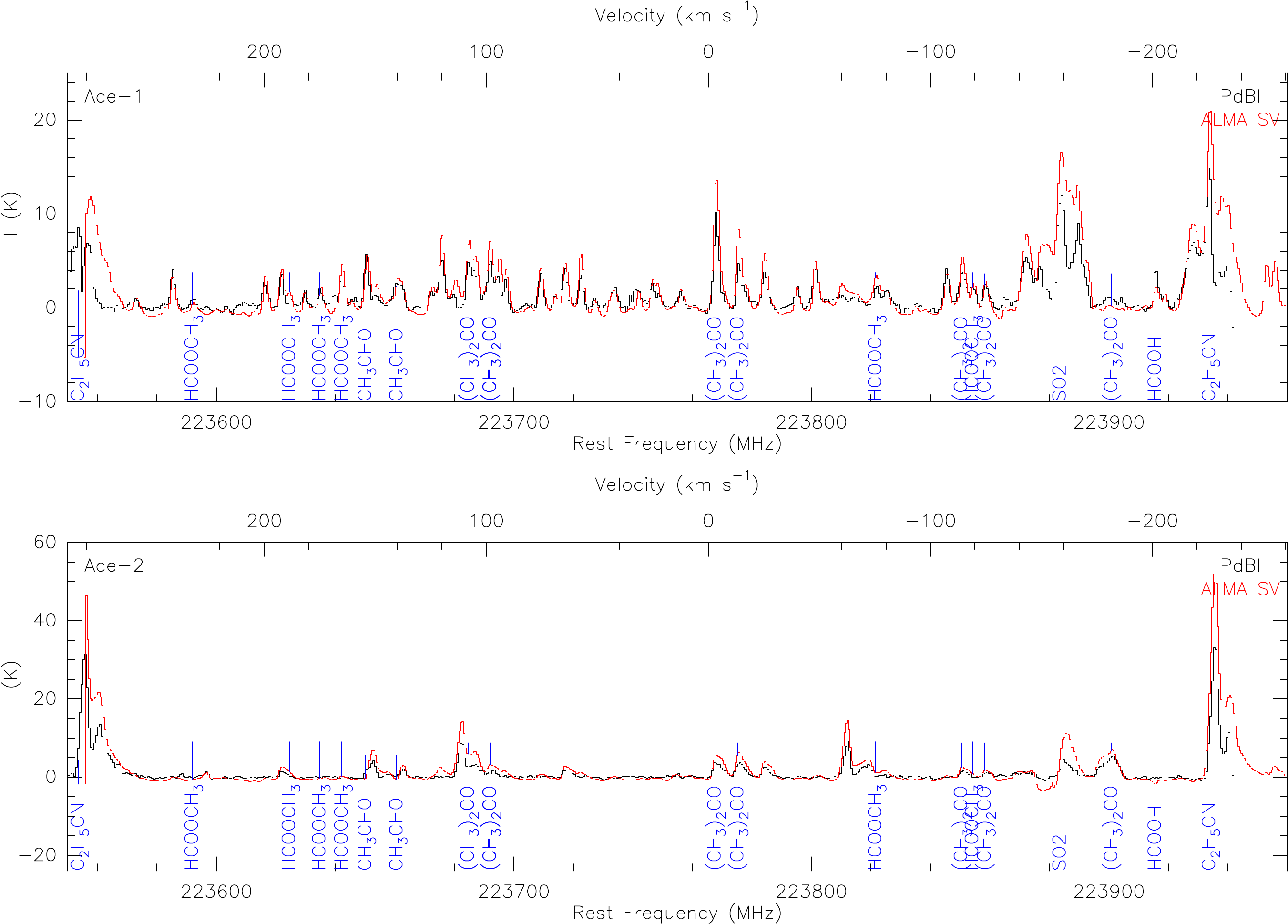}
   \caption{Comparison of the PdBI and ALMA-SV spectra near 223 GHz toward Ace-1 and Ace-2.}
   \label{pdbi-alma-spectra-1}
   \end{figure*}

\begin{figure*}
   \centering
   \includegraphics[angle=0,width=0.8\textwidth]{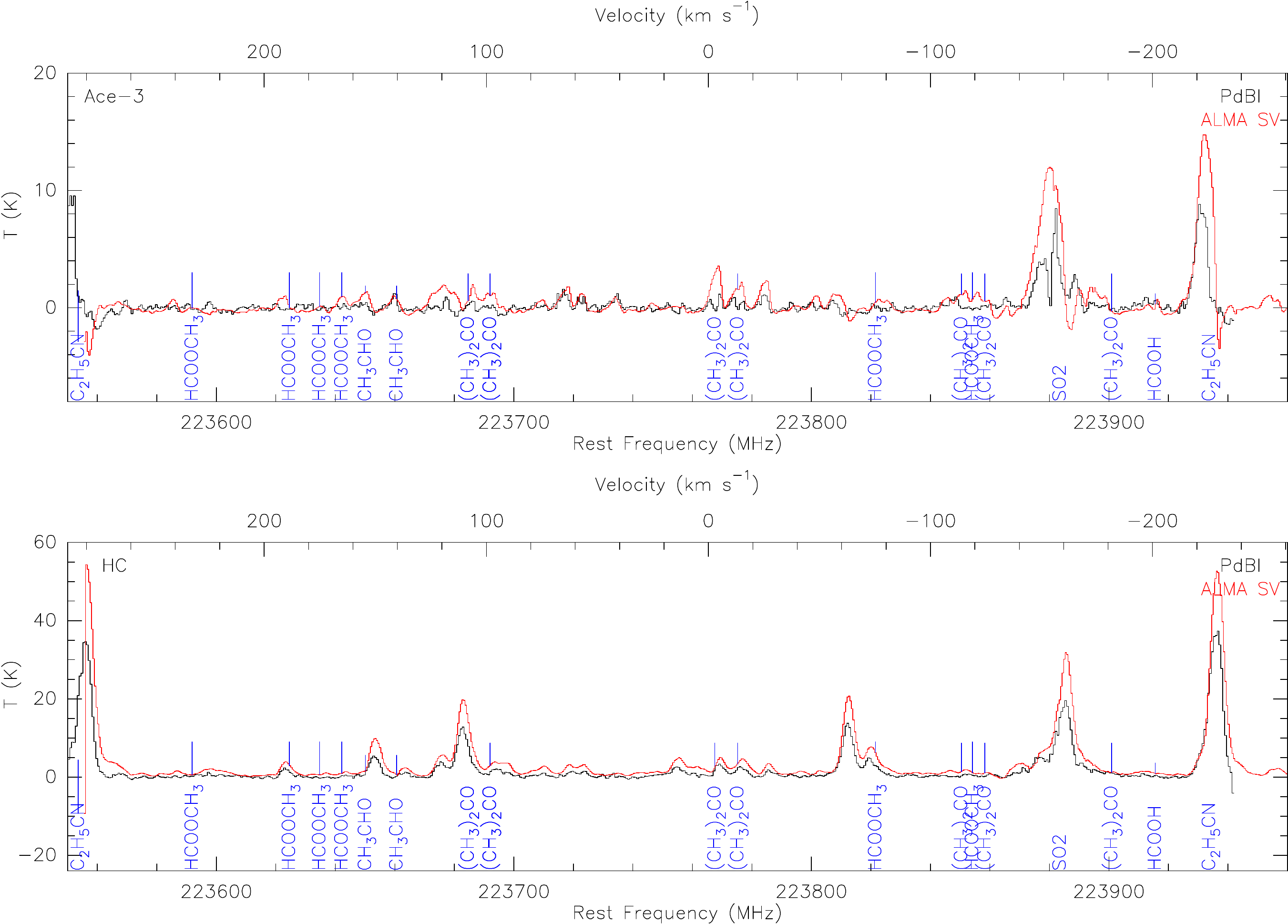}
   \caption{Comparison of the PdBI and ALMA-SV spectra near 223 GHz toward Ace-3 and HC.}
   \label{pdbi-alma-spectra-2}
   \end{figure*}

\begin{figure*}
   \centering
   \includegraphics[angle=0,width=0.8\textwidth]{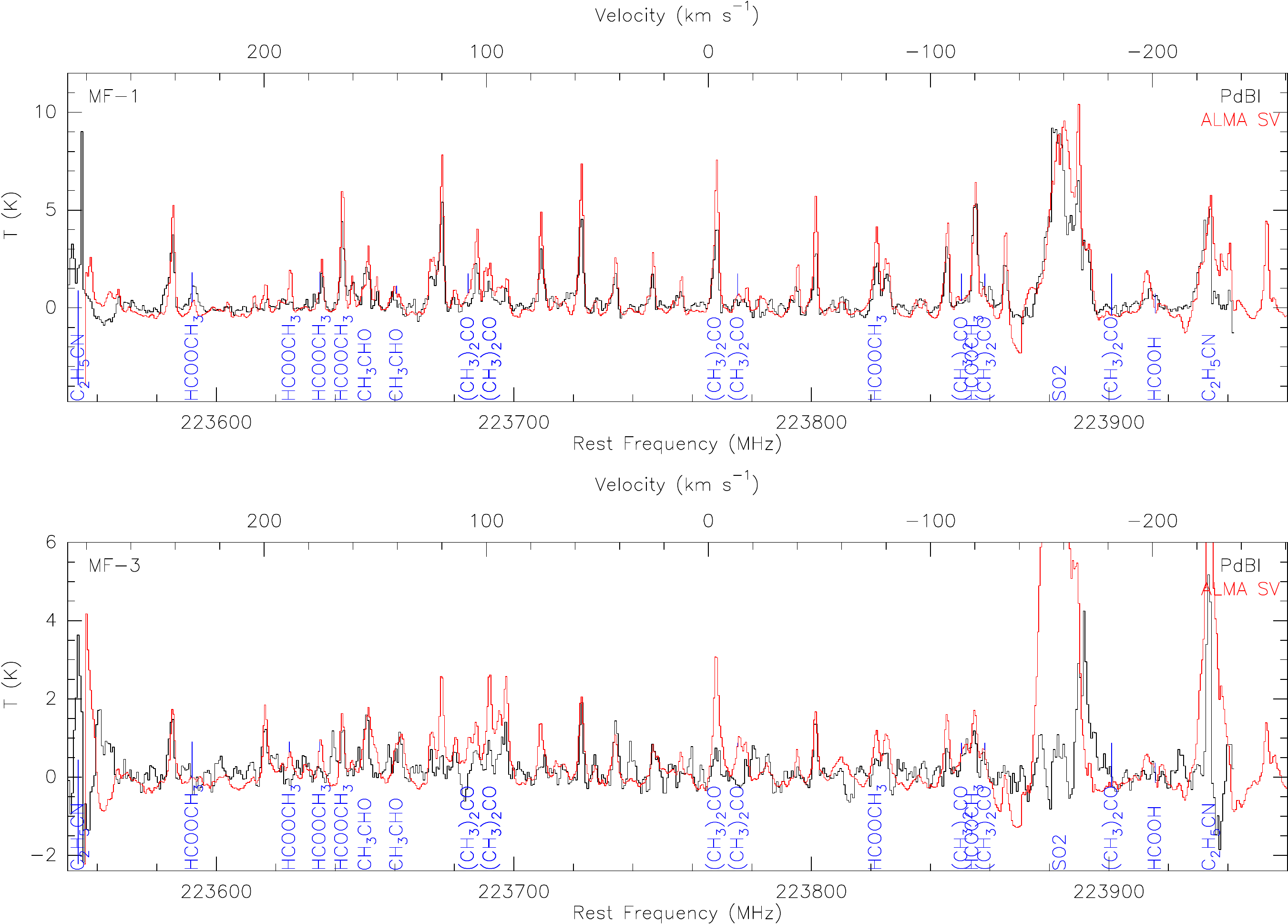}
   \caption{Comparison of the PdBI and ALMA-SV spectra near 223 GHz toward MF-1 and MF-3. MF-3, a more extended region, suffers strong filtering effects on strong broader lines, such as SO$_2$ and \ec\ in the PdBI data similar to Ace-3. Some strong lines from other molecules appear at the similar frequencies to acetone in MF-1 and MF-3.}
   \label{pdbi-alma-spectra-3}
   \end{figure*}

Figure \ref{pdbi-alma-spectra-1} shows the ALMA-SV and PdBI spectra toward Ace-1 and 2. The differences between the two data sets are small for the \mf\ and \ace\ weak lines, but greater for the \ec\ and SO$_2$ strong or broad lines. It is because these lines tend to be spatially extended and are easily resolved out by the PdBI. Filtering effects are more obvious toward Ace-3 and MF-3 (Figs. \ref{pdbi-alma-spectra-2} and \ref{pdbi-alma-spectra-3}) where most of the line emission, including SO$_2$, \ec, and \ace, is filtered out. 


\subsection*{Summary}
Our PdBI spatial distribution and individual spectra are very consistent with the new ALMA-SV data toward Orion BN/KL. However, due to different uv coverages and the smaller PdBI primary beam, some extended emission is filtered out in our data, especially toward Ace-3 and in the southern part (close to MF-3) of BN/KL. It is not clear why the arc-like acetone emission close to HC has not been resolved out. One possible reason is that the acetone emission in the southern part is weaker and suffers more easily from filtering effects, as observed toward the Ace-3 region. The fully functional ALMA in the near future with a denser uv coverage and the short spacings of the Atacama Compact Array will surely help solve these issues and greatly improve the molecular line images.

\pagebreak

\clearpage
\section{Three-dimensional view of molecular structures in BN/KL \label{app3}}

The three-dimensional visualization enables us to investigate the distributions and possible structures of different molecules in great detail. Therefore, we have prepared the three-dimensional (R.A., Dec., and \vlsr) images of different molecules in Orion BN/KL using the 3D Slicer program\footnote{http://www.slicer.org} and the complemented procedures provided by the Astronomical Medicine Project \footnote{http://am.iic.harvard.edu}. Figures \ref{3d-1}-\ref{3d-4} show the different viewing angles of the \ec, \ace, and \mf\ structures in BN/KL. In addition, the superposed images of these molecules were also plotted. In addition, a complementary 3D movie will be available on the A\&A website. 

\subsection*{\ec}

The \ec\ line has a larger line width compared with those of \ace\ and \mf\ and is clearly seen in Figures \ref{3d-3} and \ref{3d-4}. The arc-like structure close to HC is clearer in the 3D images (e.g., Figs. \ref{3d-1} and \ref{3d-2}). In addition, the slow outflow associated with IRc7 in the \amm\ emission proposed by \citet{Goddi2011b} may actually be the two different clumps in the SE and NW of IRc7 (see Figs. \ref{3d-3} and \ref{3d-4}) because of the multi-component structure in the blueshifted clump revealed in the image. This complicated blueshifted clump was detected in the methyl cyanide emission \citep{Wang2010} toward southeast of IRc7. These components may be due to the explosive outflow. However, we cannot rule out the possibility of a multi-outflow that comes from IRc7.

\subsection*{\ace}

The arc-like structure is clearly seen in the acetone 3D images. It is interesting to note that the filtered-out emission toward Ace-3 is mostly centered at 8 \kms, which creates an artificial shell-like structure in the 3D images.

\subsection*{\mf}

\begin{figure*}
   \centering
   \includegraphics[page=1,width=0.88\textwidth]{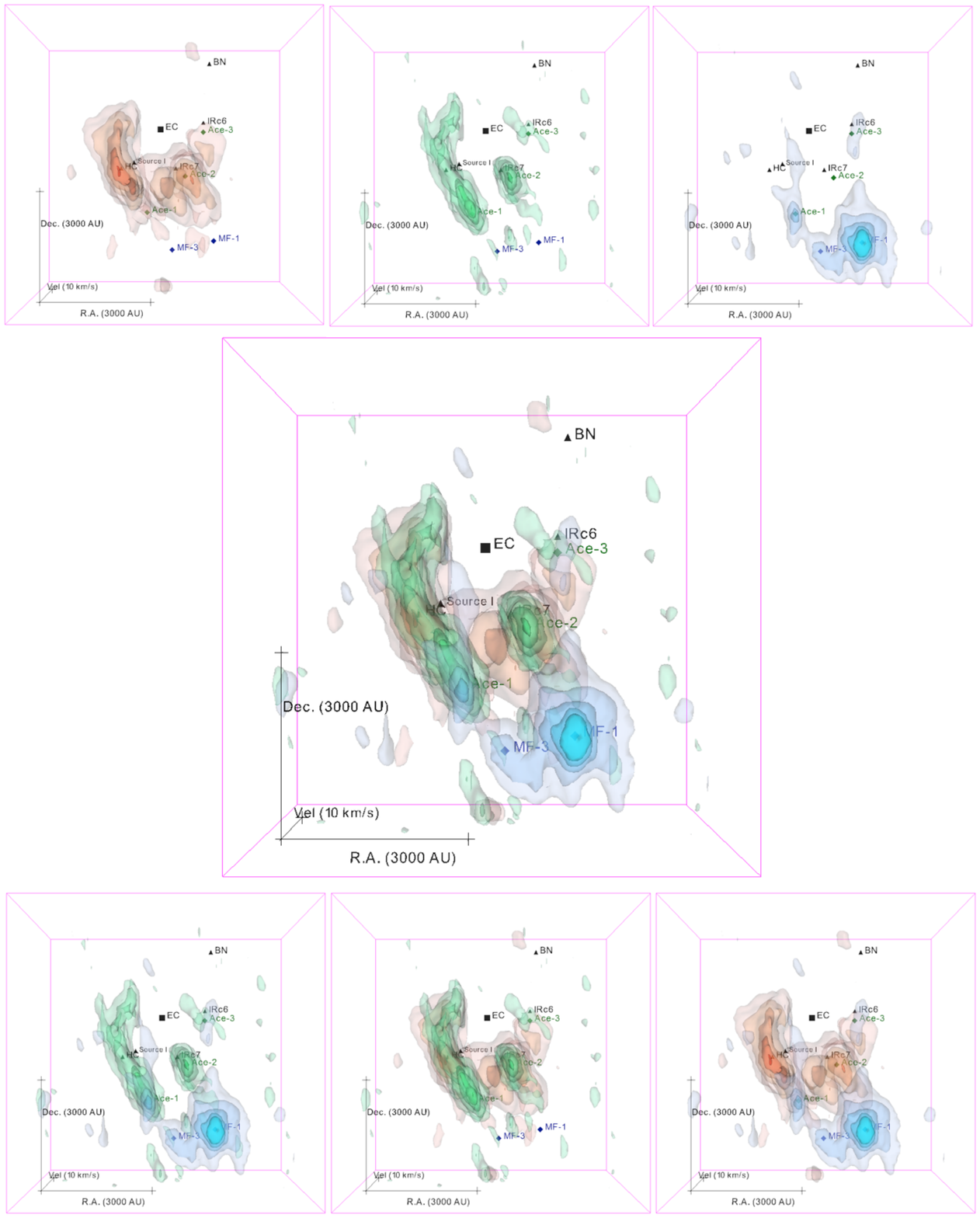}
   \caption{Three-dimensional view (R.A., Dec., and \vlsr) of \ec\ (red), \ace\ (green), and \mf\ (blue) distributions in Orion BN/KL. The \ec\ $25_{3,23}-24_{3,22}$ line at 223553.6 MHz is shown in red surfaces, the \ace\ $17_{7,11}-16_{6,10}$ EE line at 223775.3 MHz in green, and the \mf\ $11_{4,8}-10_{3,7}$ A lines at 223550.5 MHz in blue. The isothermal surfaces are plotted from 15\% to 95\% in steps of 20\% of their peak intensities.}
   \label{3d-1}
   \end{figure*}

\begin{figure*}
   \centering
   \includegraphics[page=2,width=0.88\textwidth]{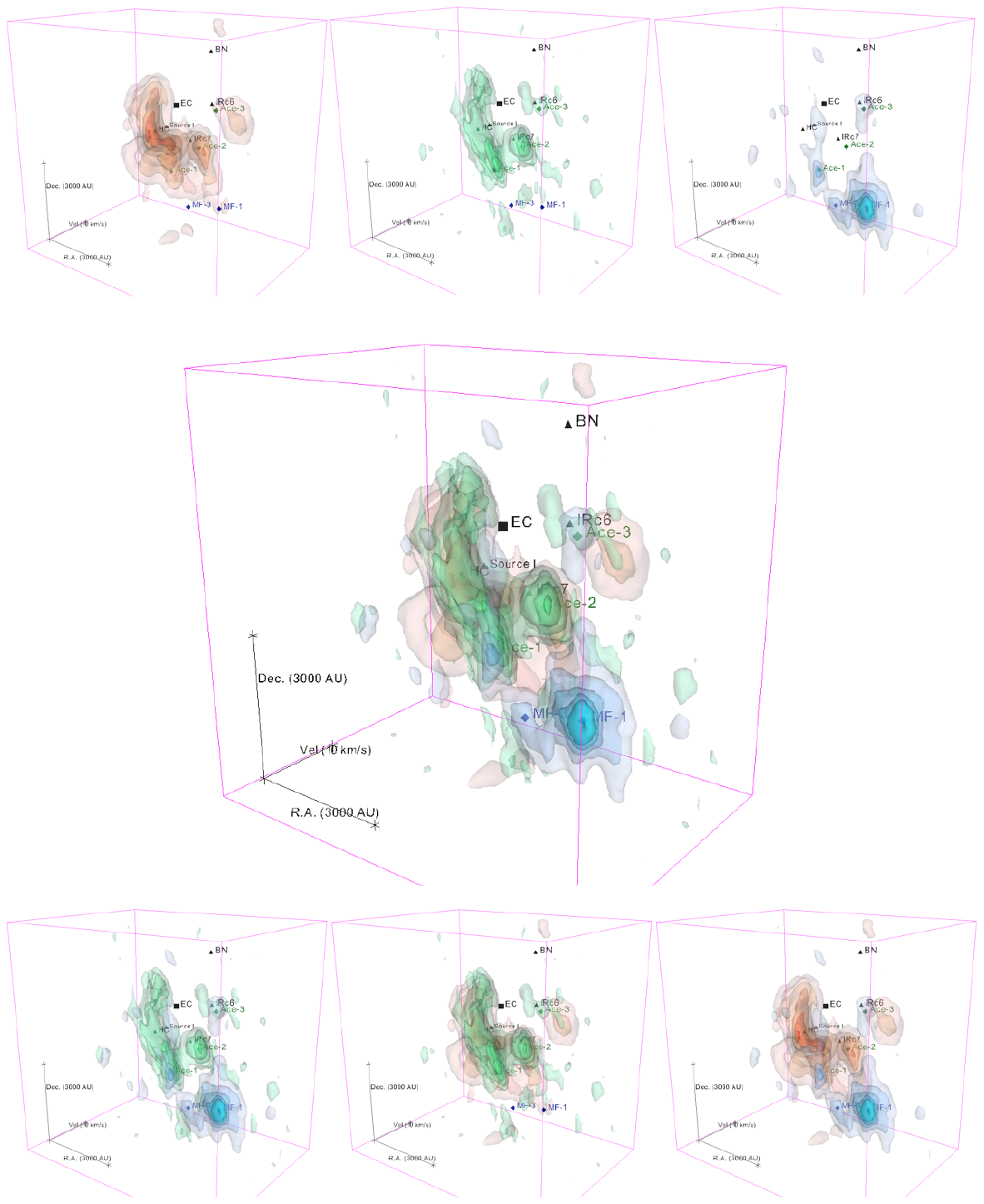}
   \caption{Three-dimensional view (R.A., Dec., and \vlsr) of \ec\ (red), \ace\ (green), and \mf\ (blue) distributions in Orion BN/KL with a different viewing angle from Fig. \ref{3d-1}.}
   \label{3d-2}
   \end{figure*}

\begin{figure*}
   \centering
   \includegraphics[page=3,width=0.88\textwidth]{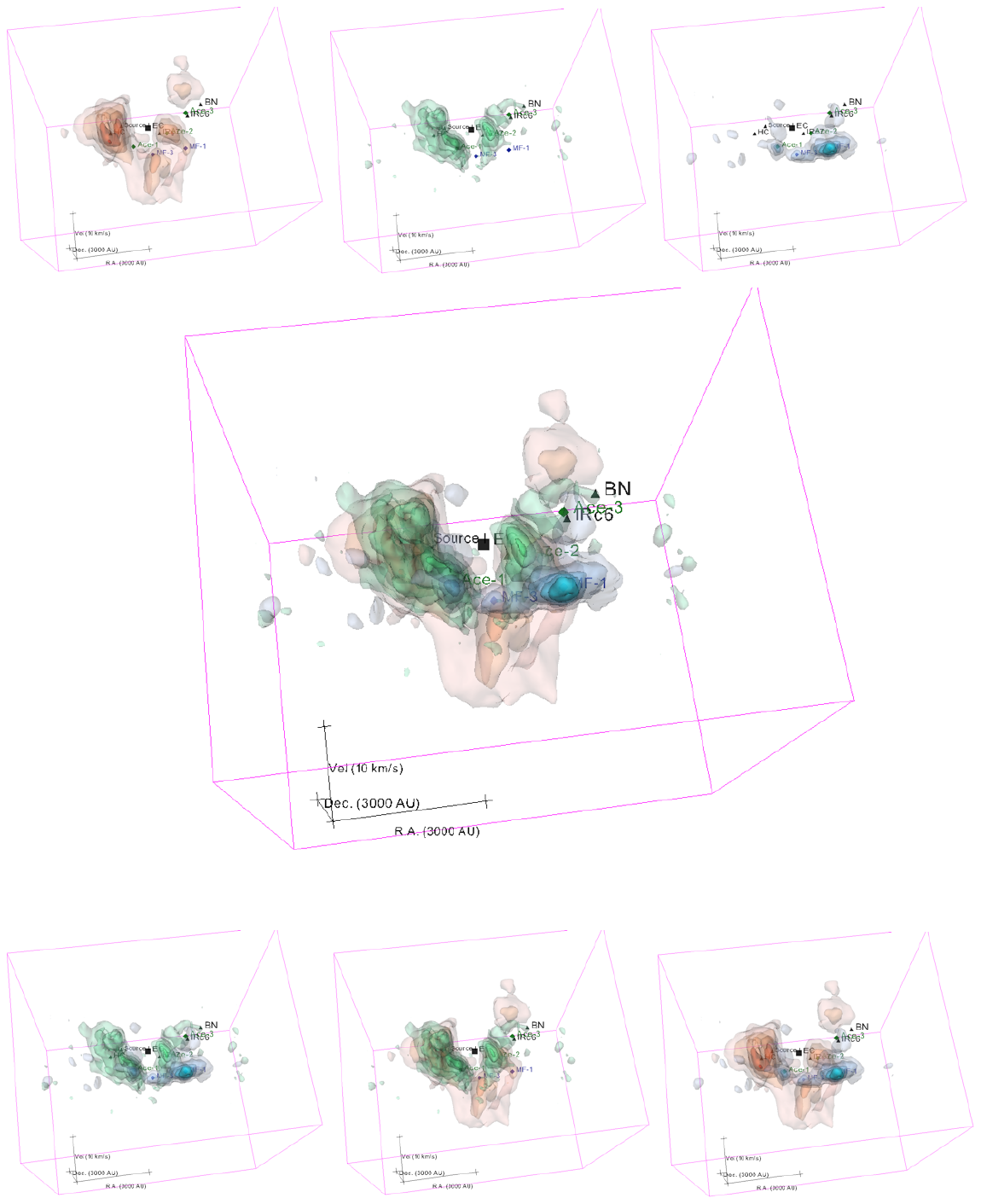}
   \caption{Three-dimensional view (R.A., Dec., and \vlsr) of \ec\ (red), \ace\ (green), and \mf\ (blue) distributions in Orion BN/KL with a different viewing angle from Fig. \ref{3d-1}.}
   \label{3d-3}
   \end{figure*}

\begin{figure*}
   \centering
   \includegraphics[page=4,width=0.88\textwidth]{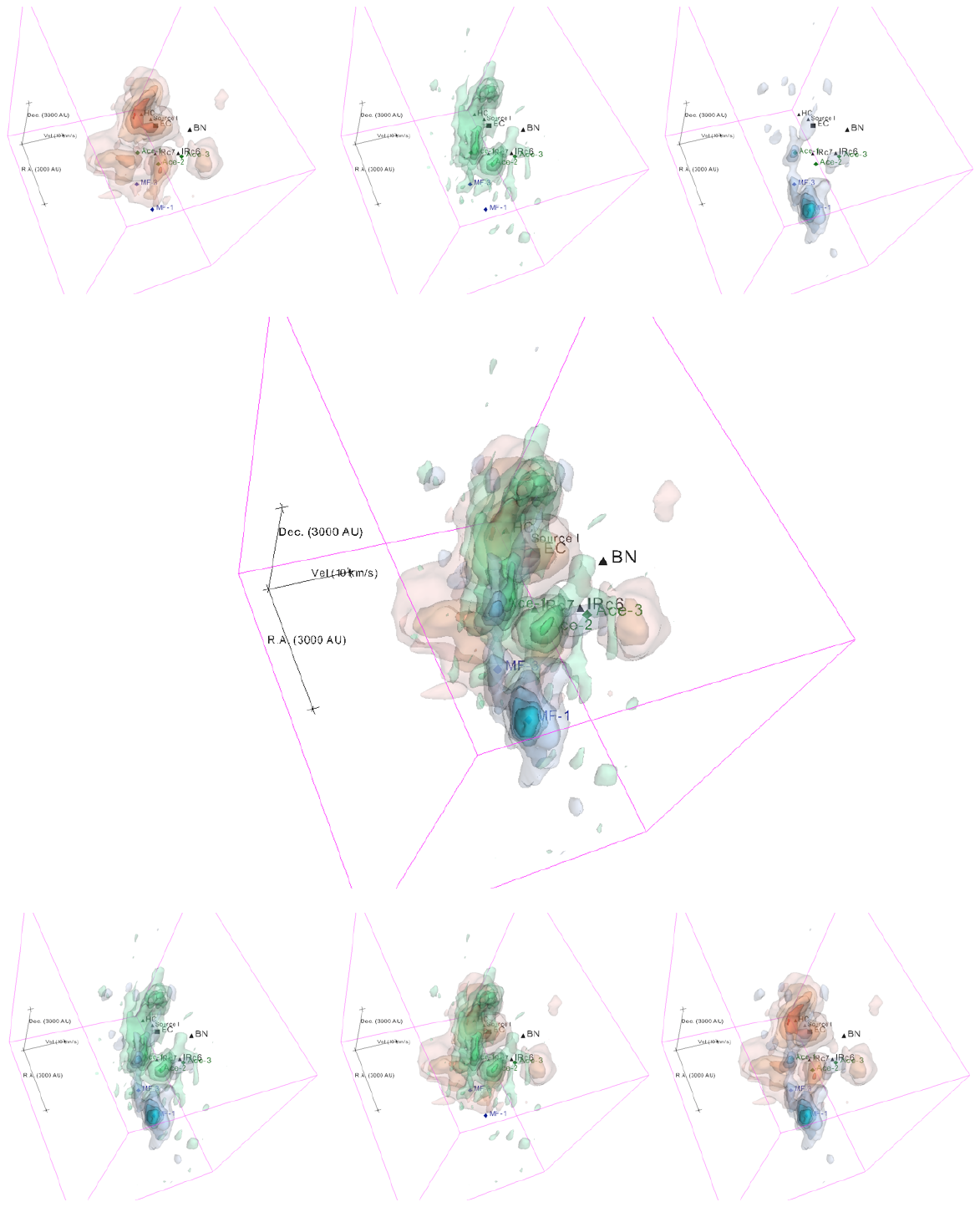}
   \caption{Three-dimensional view (R.A., Dec., and \vlsr) of \ec\ (red), \ace\ (green), and \mf\ (blue) distributions in Orion BN/KL with a different viewing angle from Fig. \ref{3d-1}.}
   \label{3d-4}
   \end{figure*}

Part of the arc-like structure seen in \ec\ and \ace\ is also seen in \mf\ where the \mf\ emission extends from MF-3 to Ace-1 and even to the northwest of HC (Figs. \ref{3d-1} and \ref{3d-2}). In addition, some structures close to MF-1 are seen and may be related to outflows/shocks.

\end{appendix}


\end{document}